\documentclass[12pt,a4]{article}

\usepackage{amsmath}

\usepackage{amssymb}

\usepackage{epsfig}

\newcommand{\pd}{\partial}
\newcommand{\NO}[1]{\,{:}#1{:}\,}
\newcommand{\com}[2]{{\big[}#1, #2{\big]}}
\newcommand{\dlt}[1]{\delta^{(#1)}(z-w)}
\newcommand{\dl}{\delta(z-w)}
\newcommand{\ee}[1]{\mathrm{e}^{\mathrm{i}#1}}
\newcommand{\ii}{\mathrm{i}}
\newcommand{\dterm}{\delta \, \text{term}}
\newcommand{\per}{\nonumber \\ &}
\newcommand{\vst}{\phantom{\rule[-0.7em]{0.1pt}{2em}}}
\newcommand{\bvect}[4]
{\begin{array}{cc}
#1
&#2
\\[4pt]
#3
&#4
\end{array}}

\newcommand{\blue}{}

\begin{document}


\begin{titlepage}

\phantom{a}

\vfill

\begin{center}
{\LARGE \bf
Quasi root systems and vertex operator
realizations of the Virasoro algebra
 \\
}
\vspace{20pt}
{
{ \large \bf Boris Noyvert
\\} 
\vspace{10pt}
{  e-mail:
\hspace{-15pt}
\raisebox{-0.7pt}{
\includegraphics[height=8pt]{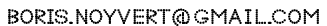}
}\\}
}

\vspace{20pt}

\begin{abstract}

{\normalsize
A construction of the Virasoro algebra
in terms of 
free massless two-dimensional boson fields is studied.
The ansatz for the Virasoro field contains the most general
unitary scaling dimension 2 expression built from vertex operators.
The ansatz leads in a natural way to a concept of
a quasi root system.
This is a new notion generalizing
the notion of a root system
in the theory of Lie algebras.
We introduce a definition of a quasi root system and provide
an extensive list of examples.
Explicit solutions of the ansatz are presented
for a range of quasi root systems.
}

\end{abstract}

\end{center}

\vfill

\end{titlepage}

\tableofcontents



\section{Introduction}

\setcounter{equation}{0}


The paper studies a bosonization of the Virasoro algebra.
The 
challenge behind this work is to provide
a unified constructive approach to the study and classification
of two-dimensional conformal field theories.
We believe that every two-dimensional conformal
field theory model can be bosonized, i.e.~represented in terms
of a number of free bosons.
Since every two-dimensional conformal model has a symmetry, described by the Virasoro algebra,
it is a natural idea to bosonize the Virasoro algebra.

There is a canonical construction of the Virasoro algebra in terms of free bosons:
\begin{equation}
                      \label{canonical}
L(z)=-\frac{1}{2}\sum_{i=1}^d 
\partial \phi_i(z) \partial \phi_i(z),
\end{equation}
where $\{\phi_i(z),i=1,\ldots,d\}$ is an orthonormal set of free boson fields.
The central charge of the Virasoro algebra in this construction
is $c=d$, the number of the free bosons.
In order to obtain noninteger values of central charge
we suggest a modified construction of the Virasoro field,
which involves vertex operators.

We introduce the following ansatz
\begin{equation}
                  \label{main_introduction}
\begin{aligned}
L(z)=&
-\frac{1}{2}\sum_{i,j=1}^d A_{i j}\,\partial \phi_i(z) \partial \phi_j(z)+
\sum_{
\left( \alpha ,\alpha \right) =4}
 b_{\alpha }\left({\mathrm{e}^{\mathrm{i}\phi_\alpha(z)}}+
{\mathrm{e}^{-\mathrm{i}\phi_\alpha(z)}}\right)
 \\
&
+\mathrm{i}\sum_{
{\left( \beta ,\beta \right) =2}}
\partial \phi_{\gamma_\beta}(z)
 \left({\mathrm{e}^{\mathrm{i}\phi_\beta(z)}}-
{\mathrm{e}^{-\mathrm{i}\phi_\beta(z)}} \right)
,
\end{aligned}
\end{equation}
and study conditions under which the field $L(z)$ satisfies
the Virasoro algebra. Here $\phi_\alpha(z)=\sum_{i=1}^d \alpha_i \phi_i$
is a free boson field associated with vector $\alpha$,
normal ordering is implied and cocycle factors
associated with the exponents of free boson fields are omitted
for the brevity of this introduction.
(The cocycles take their place a bit later in the paper.)
The expression above is a very general unitary field of
scaling dimension 2 build from free bosons.
The first term is similar to the canonical one in~(\ref{canonical}),
the difference is that a
symmetric matrix $A$ is introduced. The second term is a sum
of vertex operators of scaling dimension 2 with numerical real
coefficients $b_\alpha$. The third
term is constructed from vertex operators of scaling dimension
1 multiplied by a derivative of the free boson field.
As we will see
the ansatz generates a very rich theory and leads in a very natural way
to the introduction of a new combinatorial geometric object,
a generalization of a notion of root systems of simple Lie algebras.
We call these new generalized root systems ``quasi root systems''.

The bosonization technique was successfully employed in
different areas of mathe\-matics and physics.
We list only a few examples that are closely related to our construction.
The famous example is the bosonization
of the two-dimensional massless fermion:
\begin{equation}
\psi(z)=\frac{\mathrm{e}^{\mathrm{i}\phi(z)}
+\mathrm{e}^{-\mathrm{i}\phi(z)}}{\sqrt{2}}.
\end{equation}
The idea led to important results in condensed matter
physics and to the ``physical'' derivation of the Jacobi triple
product identity.

Another example is from the theory of affine Kac-Moody algebras.
The Frenkel-Kac construction~\cite{Frenkel:1980rn}
gives an explicit realization of a simply laced (i.e.~of ADE type)
affine Kac-Moody
algebra on level 1 in terms of a number (equal to the rank of
the algebra) of free bosons.
The Cartan generators of the affine algebra are given by
\begin{equation}
H(z)=\ii \pd \phi(z),
\end{equation}
and the generators associated to the root vectors of the algebra are expressed as
\begin{equation}
E_\alpha(z)\sim \mathrm{e}^{\mathrm{i}\phi_\alpha(z)},
\end{equation}
where the roots $\alpha$ have a standard normalization $(\alpha,\alpha)=2$.
Taking a sum of a few such constructions
one can get a realization for the affine algebra on any positive integer
level.

In the string theory oriented conformal field theory we find another
bosonization example - the Kazama-Suzuki models~\cite{Kazama:1988va}.
The authors find many new unitary models of
the $N=2$ superconformal algebra by constructing explicitly
the algebra in terms of free boson fields.
In fact the construction by Kazama and Suzuki led us to the idea of introducing
the ansatz directly for the Virasoro algebra rather than for the $N=2$ superconformal algebra.
We discuss the relation of the Kazama-Suzuki construction to our ansatz in
section~\ref{$N=2$ superconformal algebra bosonic constructions}.

The paper has the following structure.
In section~\ref{The setup} we introduce the formalism and the ingredients
for our ansatz: the free boson algebra, the language of formal distributions
and operator product expansions, and the vertex operators.
Then in section~\ref{Vertex operator realization of the Virasoro algebra}
we recall the definition of the Virasoro algebra, introduce the ansatz
and give the simplest example of the construction based just on one free boson.
In section~\ref{long formulas} we derive and tabulate
the lengthy formulae for the commutation relations
of terms forming the Virasoro field in our ansatz.
In section~\ref{Consistency conditions for the roots}
we establish the consistency conditions for the set of vectors $\alpha$ and $\beta$
in~(\ref{main_introduction})
and state our main theorem introducing quasi root systems for the first time.
The definition of quasi root systems is given in
section~\ref{Definition of quasi root system}, and
a list of examples of quasi root systems is presented in
section~\ref{Examples of quasi root systems} and in Appendix~\ref{QR_dim3}.
Section~\ref{Solutions for the ansatz} is devoted to solving the ansatz
for the range of examples of quasi root systems. We review the
literature on the subject in section~\ref{Literature review}
outlining the differences between the constructions studied there and our ansatz.
In section~\ref{Open problems} we 
discuss possible future developments in the subject.
Appendix~\ref{More general constructions} concerns with
possible generalizations of our ansatz.


\section{The setup}

\label{The setup}

\setcounter{equation}{0}


In this section we briefly introduce the algebraic apparatus
needed for the bosonization construction.


\subsection{Free boson system}


The starting point is the Heisenberg algebra,
which is probably the simplest infinite dimensional Lie algebra.
It is generated by
$a_n, n\in \mathbb{Z}$ and an additional generator $q$,
subject to the following commutation relations:
\begin{equation}
\begin{aligned}
\com{a_n}{a_m}&
=n \, \delta_{m+n,0}\, ,
\\
\com{q}{a_n}&=\ii \, \delta_{n,0},
\end{aligned}
 \quad n,m \in \mathbb{Z}.
\end{equation}
In two-dimensional conformal field theory this algebra
describes a free boson system.

We introduce formal power series
(we'll call them ``fields'' following the physical language)
with algebra generators  serving as coefficients
of the series. So we define a free boson field $\phi(z)$
and its derivative $\partial \phi(z)$:
\begin{align}
\phi(z)&=\ii \sum_{n\in\mathbb{Z} \atop n \neq 0 } 
\frac{a_n}{n} z^{-n}-\ii \, a_0 \ln{z}+q,\\
\partial \phi(z)&
=-\ii \sum_{n\in\mathbb{Z}} a_n z^{-n-1}.
\end{align}
It is convenient to split the free boson field into ``positive''
and ``negative'' parts:
$\phi(z)=\phi^-(z)+\phi^+(z)$,
where
\begin{align}
\phi^-(z)&=\ii\sum_{ n < 0 }
\frac{a_n}{n} z^{-n}+q,\\
\phi^+(z)&=\ii \sum_{ n > 0 }
\frac{a_n}{n} z^{-n}-\ii a_0 \ln{z}.
\end{align}
We essentially also split the generators of the Heisenberg algebra
into two groups, creation ($q, a_n, n < 0$) and annihilation
($ a_n, n \ge 0$) operators.
Note that the creation operators are commutative between themselves,
as well as the annihilation operators. And therefore
$$
\com{\phi^+(z)}{ \phi^+(w)}=\com{\phi^-(z)}{ \phi^-(w)}=0.
$$

At this point we can introduce the
normal ordering convention.
Normal ordering $\NO{\phantom{a}}$ of a product of Heisenberg algebra generators
means that the generators are simply reordered so that
all the creation operators are to the left from the annihilation
operators. For example
$\NO{a_2\, a_0 \,q \,a_{-2}}=a_{-2} \, q \, a_0 \,a_{2}$.

Next we calculate the commutator between the positive and negative
parts of the free boson field:
\begin{equation}
\com{\ii \phi^+(z)}{\ii \phi^-(w)}=-\sum_{ n > 0 }
\frac{1}{n}\left(\frac{w}{z}\right)^n + \ln{z}
\end{equation}
The series here is an expansion of a function $\ln{(1-w/z)}$
in $w$ around $0$ in the region $|z|>|w|$. We will denote it by
the following:
\begin{equation}
s_{z>w} \ln{(1-w/z)}=-\sum_{ n > 0 }
\frac{1}{n}\left(\frac{w}{z}\right)^n
\end{equation}
And we will also use the similar notation for other
series:
\begin{align}
s_{z>w} \frac{1}{z-w}=\frac{1}{z}\sum_{ n \ge 0 }
\left(\frac{w}{z}\right)^n,\\
s_{z>w} \frac{1}{(z-w)^s}=\frac{1}{z^s}\sum_{ n \ge 0 }
{n+s-1 \choose s-1}
\left(\frac{w}{z}\right)^n .
\end{align}
So finally one can write
\begin{equation}
                      \label{phi_commutator}
\com{\ii \phi^+(z)}{\ii \phi^-(w)}=s_{z>w} \ln{(z-w)},
\end{equation}

The theory is easily generalized to the case of $d$-dimensional
free boson fields. One just takes $d$ mutually commutative copies
of the Heisenberg algebra, which form $d$ independent bosonic fields
$\phi_\mu(z), \ \mu=1,\ldots,d$. The commutation
relation~(\ref{phi_commutator}) in the multidimensional form reads
\begin{equation}
                      \label{phi_commutator_multi}
\com{\ii \phi^+_\mu(z)}{\ii \phi^-_\nu(w)}=\delta_{\mu\, \nu}s_{z>w}\ln{(z-w)},
\qquad
\mu,\nu=1,\ldots,d.
\end{equation}
We embed these free boson fields into the
$d$-dimensional Euclidean space $\mathbb{E}^d$ with a standard
scalar product, which we denote by $(\cdot,\cdot)$.
The notation $\phi_\alpha (z), \alpha \in \mathbb{E}^d$
will stand for
\begin{equation}
\phi_\alpha (z)=\sum_{\mu=1}^d \alpha_\mu \phi_\mu(z),
\end{equation}
where $\alpha_\mu$ are the Euclidean components of vector $\alpha$.
In this notation the analogue of~(\ref{phi_commutator_multi})
reads
\begin{equation}
                      \label{phi_commutator_alpha}
\com{\ii \phi^+_\alpha(z)}{\ii \phi^-_\beta(w)}=(\alpha,\beta)s_{z>w}\ln{(z-w)}.
\end{equation}
And there is a similar formula for derivatives of boson fields:
\begin{equation}
                      \label{pd_phi_commutator_alpha}
\com{\pd \phi^+_\alpha(z)}{\pd \phi^-_\beta(w)}=-(\alpha,\beta)s_{z>w}\frac{1}{(z-w)^2}\, .
\end{equation}

Now we introduce the exponent of bosonic fields by the standard
Taylor series:
\begin{equation}
\ee{\phi_\alpha(z)}=\sum_{k=0}^\infty
\frac{1}{k!} (\ii \phi_\alpha(z))^k.
\end{equation}
We will only deal with the normal ordered exponents,
which are well defined operators from the physical point of view:
\begin{equation}
\NO{\ee{\phi_\alpha(z)}}=\sum_{k=0}^\infty
\frac{1}{k!} \ii^k \NO{(\phi_\alpha(z))^k}.
\end{equation}
A very important fact is that this ordered exponent can be written as
\begin{equation}
\NO{\ee{\phi_\alpha(z)}}=\ee{\phi^-_\alpha(z)}\ee{\phi^+_\alpha(z)}.
\end{equation}
This identity is easily proved using the following observation
\begin{equation}
\NO{(\phi_\alpha(z))^k}=\sum_{j=0}^k
\left({k \atop j}\right)
({\phi^-_\alpha(z)})^j ({\phi^+_\alpha(z)})^{k-j},
\end{equation}
which is true due to the normal ordering on the left hand side.

The following formulae are the key for our future calculations:
\begin{align}
                    \label{phi+exp}
\com{\pd \phi^+_\alpha(z)}{\ee{\phi^-_\beta(w)}}&=
-s_{z>w}\, \ii\frac{(\alpha,\beta)}{z-w}{\ee{\phi^-_\beta(w)}},\\
                    \label{exp_phi-}
\com{\ee{\phi^+_\alpha(z)}}{\pd \phi^-_\beta(w)}
&=
s_{z>w}\, \ii\frac{(\alpha,\beta)}{z-w}{\ee{\phi^+_\alpha(z)}},\\
                    \label{exp+exp-}
{\ee{\phi^+_\alpha(z)}}{\ee{\phi^-_\beta(w)}}&=s_{z>w}(z-w)^{(\alpha,\beta)}
{\ee{\phi^-_\beta(w)}}{\ee{\phi^+_\alpha(z)}}.
\end{align}
The last identity is proven using the Baker-Campbell-Hausdorff formula:
\begin{equation}
\mathrm{e}^X \mathrm{e}^Y= \mathrm{e}^Y \mathrm{e}^X \mathrm{e}^{[X,Y]},
\qquad
\text{if } [X,[X,Y]]=[Y,[X,Y]]=0.
\end{equation}

The identities
(\ref{pd_phi_commutator_alpha}, \ref{phi+exp},
\ref{exp_phi-}, \ref{exp+exp-})
are useful for reordering the fields to the normal order
in the computation
of commutation relations in Section~\ref{long formulas}.


\subsection{Delta-function formalism}


Here we
define the
notion of a formal delta-function,
the definition is taken from the book by Kac~\cite{Kac book}.
A formal delta-function $\dl$ is given by the following
formal power series:
\begin{equation}
 \dl=z^{-1}\sum_{n\in\mathbb{Z}}(w/z)^n.
\end{equation}
 It is easy to see that
\begin{equation}
                 \label{delta_def}
 \dl=s_{z>w}\frac{1}{z-w}-s_{w>z}\frac{1}{z-w}.
\end{equation}
We will also need the derivatives of the formal delta-function:
\begin{equation}
                 \label{delta_der_def}
\delta^{(s)}(z-w)=\frac{1}{s!}\partial_w^s \delta(z-w)=
s_{z>w}\frac{1}{(z-w)^{s+1}}-s_{w>z}\frac{1}{(z-w)^{s+1}}
\end{equation}
The standard rules for the Dirac delta-function apply,
the most important one is
\begin{align}
(z-w)\,\dlt{k}=\dlt{k-1}, \qquad k>0.
\end{align}
The delta-function language is useful in computations of different
commutation relations. For example the commutator of two derivatives
of the free boson field, obtained from~(\ref{pd_phi_commutator_alpha}),
can be written as
\begin{equation}
                      \label{bosons_commutation}
\com{\pd \phi_\alpha(z)}{\pd \phi_\beta(w)}=-(\alpha,\beta)\dlt{1}\, .
\end{equation}

In what follows the commutators of fields will be always expressed in terms
of the formal delta function and its derivatives. Readers
who are more familiar with the language of operator product expansions
commonly used in physical literature on conformal field theory
can just simply replace the commutators by operator product expansions
and the delta functions by negative powers of $(z-w)$ following the rule:
\begin{equation}
\dlt{n}\to\frac{1}{(z-w)^{n+1}}.
\end{equation}
Then for example the analogue of the commutator
relation~(\ref{bosons_commutation}) will look like
\begin{equation}
{\pd \phi_\alpha(z)}{\pd \phi_\beta(w)}=-\frac{(\alpha,\beta)}{(z-w)^2}
+ \text{regular part}
\end{equation}
in the formalism of operator product expansions.


\subsection{Vertex operators}

\label{Vertex operators}


In this section we again follow the lines of the book~\cite{Kac book}.

Vertex operators
\begin{equation}
\Gamma_\alpha(z)=c_\alpha \, \NO{\ee{\phi_\alpha(z)}}
\end{equation}
are defined as bosonic exponents multiplied by cocycles $c_\alpha$,
which are
elements of a twisted group algebra,
here $\alpha \in \mathbb{E}^d$.
The twisted group algebra  is well defined in the case
of integer lattice only. So we take
$\alpha \in Q \subset \mathbb{E}^d$,
where $Q$ is an integer lattice.
The elements $c_\alpha$ satisfy
a twisted group algebra $\mathbb{C}_\epsilon[Q]$
with relation
\begin{equation}
                  \label{cocycle_product}
c_\alpha c_\beta=\epsilon(\alpha,\beta)c_{\alpha+\beta},
\end{equation}
where $\epsilon(\alpha,\beta)$ is a 2-cocycle.
The 2-cocycle is defined by the following properties
\begin{align}
                      \label{cocycle_1}
&\epsilon(0,\alpha)=\epsilon(\alpha,0)=1,\\
                      \label{cocycle_2}
&\epsilon(\alpha,\beta) \epsilon(\alpha+\beta,\gamma)=
\epsilon(\alpha,\beta+\gamma) \epsilon(\beta,\gamma),\\
                      \label{cocycle_3}
&\epsilon(\alpha,\beta)=
(-1)^{(\alpha,\beta)+(\alpha,\alpha)(\beta,\beta)}\epsilon(\beta,\alpha),
\end{align}
here all $\alpha,\beta,\gamma \in Q$.
One can always find a cocycle
which is bimultiplicative
\begin{equation}
                       \label{bimultiplicativity}
\begin{aligned}
\epsilon(\alpha,\beta+\gamma)&=
\epsilon(\alpha,\beta)\epsilon(\alpha,\gamma),\\
\epsilon(\beta+\gamma,\alpha)&=
\epsilon(\beta,\alpha)\epsilon(\gamma,\alpha)
\end{aligned}
\end{equation}
and takes values $\epsilon(\alpha,\beta)=\pm 1$
for all $\alpha,\beta \in Q$.
The proof is given by the following explicit procedure.
Choose an ordered basis of $Q$: $\alpha_1, \ldots, \alpha_d$,
define the 2-cocycle on the basis vectors:
\begin{equation}
\begin{aligned}
\epsilon(\alpha_i,\alpha_j)=
\left\{
\begin{array}{ll}
1, & i<j,\\
(-1)^{\frac{1}{2}(\alpha_i,\alpha_i)((\alpha_i,\alpha_i)+1)},
& i=j,\\
(-1)^{(\alpha_i,\alpha_j)+(\alpha_i,\alpha_i)(\alpha_j,\alpha_j)},
& i>j,
\end{array}
\right.
\end{aligned}
\end{equation}
and extend the 2-cocycle to all the vectors on the lattice $Q$ by
bimultiplicativity.

It is easy to prove that in the above choice of cocycle
the following identity takes place
\begin{equation}
\epsilon(\alpha,\alpha)=
(-1)^{\frac{1}{2}(\alpha,\alpha)((\alpha,\alpha)+1)}
\end{equation}
for any vector $\alpha \in Q$. In particular
$\epsilon(\alpha,\alpha)=1$, if $(\alpha,\alpha)=4$,
and $\epsilon(\beta,\beta)=-1$, if $(\beta,\beta)=2$.

It follows from~(\ref{cocycle_1}) and (\ref{bimultiplicativity})
that $\epsilon(\alpha,-\beta)=\epsilon(\alpha,\beta)
=\epsilon(-\alpha,\beta)=\epsilon(-\alpha,-\beta)$.


\section{Vertex operator realization of the Virasoro algebra}

\label{Vertex operator realization of the Virasoro algebra}

\setcounter{equation}{0}


The Virasoro algebra is the most important algebra in conformal field
theory. It is an infinite dimensional Lie algebra generated
by $L_n, n\in \mathbb{Z}$ and $C$, subject to the following
commutation relations:
\begin{align}
\com{L_n}{L_m}&=(n-m)L_{n+m}+ C\,\frac{n(n^2-1)}{12}\delta_{n+m,0}\, ,\\
\com{C}{L_n}&=0.
\end{align}
$C$ is the center of the algebra, it can always be set
proportional to identity:
$C=c \, \mathrm{I}$.
The complex number $c$ is called the central charge.

Here we also switch to the language of formal power series
by introducing the 
Virasoro field:
\begin{equation}
L(z)=\sum_{n \in \mathbb{Z}} L_n z^{-n-2}
\end{equation}
Then the Virasoro commutation relation in terms of fields reads
\begin{equation}
              \label{Virasoro relation}
\com{L(z)}{L(w)}=\frac{c}{2}\dlt{3}+2L(w)\dlt{1}+\pd L(w)\dl .
\end{equation}

We are going to construct a general realization
of the Virasoro algebra
in terms of free bosons.
There is a well known canonical form
of the Virasoro algebra associated to the free boson system:
\begin{equation}
                 \label{canonical construction}
L^\text{st}(z)=-\frac{1}{2}\sum_{\mu=1}^d{:}\partial \phi_\mu(z) \partial \phi_\mu(z){:},
\end{equation}
where $\{\phi_\mu, \mu=1,\ldots,n\}$ is an orthonormal basis of $\mathbb{E}^n$. The field
$L^\text{st}(z)$
generates the Virasoro algebra with central charge $c=d$.
We are looking for nonstandard 
realizations of the Virasoro algebra involving vertex operators.
So we introduce the following ansatz:
\begin{equation}
                  \label{main}
\begin{aligned}
L(z)=&
-\frac{1}{2}\sum_{\mu,\nu=1}^d A_{\mu\nu}\,{:}\partial \phi_\mu(z) \partial \phi_\nu(z){:}+
\sum_{
\left( \alpha ,\alpha \right) =4}
 b_{\alpha }\left(\Gamma _{\alpha}(z)+
 \Gamma _{-\alpha }(z)\right)
 \\
&
+\mathrm{i}\sum_{
{\left( \beta ,\beta \right) =2}}
 {:}
\partial \phi_{\gamma_\beta}(z)
 \left(\Gamma _{\beta }(z)- \Gamma _{-\beta }(z)\right){:},
\end{aligned}
\end{equation}
here $\alpha,\beta, \gamma_\beta \in \mathbb{E}^n$, $(\gamma_\beta,\beta)=0$,
$A_{\mu\nu}$ is a symmetric real matrix,
$b_\alpha \in \mathbb{R}$. 

This is a very general form of the Virasoro field, which is apriori unitary.
The first term in~(\ref{main}) is similar to the standard construction
in~(\ref{canonical construction}), one just introduces the matrix $A$ into the term.
The second term is a sum of vertex operators based on vectors $\alpha$
of square length 4. We will call such vectors ``long roots''. The third term
is composed of vertex operators based on vectors $\beta$ of square length 2,
multiplied by a derivative of the boson field. We will refer to such vectors
$\beta$ as ``short roots''.

One still has to check that the expression in~(\ref{main})
indeed satisfies the Virasoro algebra. There will be additional
constraints on the roots and coefficients.
We will see later that the problem of finding a solution
to the Virasoro algebra commutation relation in form~(\ref{main})
can be divided to two parts: first -- find the appropriate set of roots,
second -- solve a set of quadratic equations for the coefficients
$A_{\mu\nu}$, $b_{\alpha}$, $\gamma_\beta$.

Now we would like to illustrate the idea on a simple example.
The simplest example involves just one free boson
field $\phi(z)$. In one dimension 
the ansatz~(\ref{main}) becomes
\begin{equation}
                         \label{main_in_one_dimension}
L(z)=
-\frac{a}{2}\,{:}\partial \phi(z) \partial \phi(z){:}+
b
\left(\mathrm{e}^{2\mathrm{i}\phi(z)}+
\mathrm{e}^{-2\mathrm{i}\phi(z)}\right),
\end{equation}
We omit the cocycles since in the case of one-dimensional
even lattice the twisted group algebra is trivial
(i.e.~all 2-cocycles $\epsilon(\alpha,\beta)$ can be set to one).
We have to check that the expression satisfies
the Virasoro algebra
commutation relation~(\ref{Virasoro relation}).
We do not want to present here the explicit calculation
of the commutator, since we do it later in the general case.
But readers familiar with the basics of conformal field theory
should be able to do this easy calculation.
The result is that the field~(\ref{main_in_one_dimension})
generates the Virasoro algebra provided
the following
set of equations is satisfied:
\begin{equation}
\begin{aligned}
a&=a^2+4b^2, \\
b&=2 a b. 
\end{aligned}
\end{equation}
Then the central charge $c$ is given by
\begin{equation}
c=a^2+4b^2.
\end{equation}
The set of equations has 4 solutions:
\begin{equation}
\begin{array}{lll}
\text{trivial:} &a=b=0 &(c=0), \\
\text{standard:} &a=1,\, b=0 & (c=1),\\
\text{nonstandard:} &a=1/2,\, b=\pm 1/4  &(c=1/2) .
\end{array}
\end{equation}
The two nonstandard solutions represent the $c=1/2$ minimal model
of the Virasoro algebra (the Ising model, see ref.~\cite{Belavin:1984vu}).

The expression in our ansatz~(\ref{main}) is not the most general
unitary field of scaling dimension 2 built from vertex operators and free
boson fields. We discuss a more general expression
in Appendix~\ref{More general constructions} and argue that all the interesting
cases are essentially covered by our ansatz.


\section{Commutator formulae for vertex operators}

\label{long formulas}

\setcounter{equation}{0}


In this section we want to compute explicit formulae
for commutators of the terms entering into the expression
in~(\ref{main}). Our aim is to list all
the mutual commutation relations between the following
three types of terms:
\begin{align}
\Omega_A(z)&=-\frac{1}{2} A_{\mu \nu} \NO{\pd \phi_\mu(z)\pd \phi_\nu(z)},\\
\Gamma_\alpha(z)&=c_\alpha \, \NO{\ee{\phi_\alpha(z)}},\\
\Lambda_{\beta,\gamma}(z)&=c_\beta \,
\NO{\pd \phi_\gamma(z)\ee{\phi_\beta(z)}},
\quad
(\gamma,\beta)=0,
\end{align}
the matrix $A_{\mu \nu}$ is always assumed to be symmetric.

To calculate a commutator $\com{F(z)}{G(w)}$ of two expressions
involving boson fields one has to reorder the boson fields
in ${F(z)}{G(w)}$ and in ${G(w)}{F(z)}$ so that all $\phi^+$
are to the right and all $\phi^-$ are to the left.
Then the normal ordered parts cancel out:
$\NO{F(z)G(w)}-\NO{G(w)F(z)}=0$ and one should stay with a number
of terms proportional to the delta-function and its derivatives.

In our computations we will keep the $\dlt{n},\, n\ge 1$ terms
and usually skip the $\dl$ term. One doesn't need the $\dl$ term
since in the Virasoro algebra commutation relation~(\ref{Virasoro relation})
the $\pd L(w) \dl$ term is satisfied automatically provided
the more singular terms are satisfied.

We start from the commutators of the $\Omega_A(z)$ term.
The result of the computation is
\begin{align}
                  \label{Omega1}
\com{\Omega_A(z)}{\Omega_A(w)}&=\frac{1}{2} \dlt{3}\mathrm{Tr}(A^2)
-\dlt{1}A_{\mu \nu} A_{\mu \rho}\NO{\pd \phi_\nu(w)\pd \phi_\rho(w)} - \per
-\dl A_{\mu \nu} A_{\mu \rho}\NO{\pd^2 \phi_\nu(w)\pd \phi_\rho(w)},\\
                  \label{Omega2}
\com{\Omega_A(z)}{\Gamma_\alpha(w)}&=\dlt{1}\frac{(\alpha, A \alpha)}{2}{\Gamma_\alpha(w)}
+\dterm ,\\
                  \label{Omega3}
\com{\Omega_A(z)}{\Lambda_{\alpha,\gamma}(w)}&=
-\ii \dlt{2}(\alpha,A \gamma)\Gamma_\alpha(w)+ \per
+\dlt{1}\left(
\frac{(\alpha, A \alpha)}{2}\Lambda_{\alpha,\gamma}(w)+
\Lambda_{\alpha,A \gamma}(w)
\right)+
\dterm ,
\end{align}
where ``$\dterm$'' denotes the term proportional to $\dl$,
and summation is implied on repeated indices.
$A \alpha$ stands for a product of matrix $A$ and vector $\alpha$:
$(A \alpha)_\mu=\sum_{\nu=1}^d A_{\mu \nu}\alpha_\nu.$

The rest of the commutators is given below. 
For these commutators we have to assume that $(\alpha,\beta)\in \mathbb{Z}$,
$(\alpha,\alpha)\in \mathbb{Z}$, $(\beta,\beta)\in \mathbb{Z}$
and $(\alpha,\alpha) (\beta,\beta)\in 2\mathbb{Z}$ (i.e.~either
 $(\alpha,\alpha)$ or $(\beta,\beta)$ or both are even):
\begin{align}
                                       \label{Gamma-Gamma}
\com{\Gamma_\alpha(z)}{\Gamma_\beta(w)}&=
\left\{
\begin{array}{ll}
0, &(\alpha,\beta)\ge 0,\\
\dlt{-(\alpha,\beta)-1}\Gamma_{\alpha,\beta}(z,w)\epsilon(\alpha,\beta),
&(\alpha,\beta)< 0,
\end{array}
\right.\\
                             \label{Gamma-Lambda}
\com{\Gamma_\alpha(z)}{\Lambda_{\beta,\gamma}(w)}&=
\ii(\alpha,\gamma) \dlt{-(\alpha,\beta)} \Gamma_{\alpha,\beta}(z,w) \epsilon(\alpha,\beta) + \per +
\dlt{-(\alpha,\beta)-1}
\NO{\Gamma_{\alpha,\beta}(z,w) \pd \phi_\gamma(w)}
\epsilon(\alpha,\beta),
\\
                            \label{Lambda-Lambda}
\com{\Lambda_{\alpha,\rho}(z)}{\Lambda_{\beta,\gamma}(w)}
&=\dlt{-(\alpha,\beta)-1}
\NO{\pd \phi_\rho(z) \Gamma_{\alpha,\beta}(z,w) \pd \phi_\gamma(w)}
\epsilon(\alpha,\beta)+ \per
+\ii \dlt{-(\alpha,\beta)}  \bigg(
(\alpha,\gamma)\NO{\pd \phi_\rho(z) \Gamma_{\alpha,\beta}(z,w)}- \per-
(\beta,\rho)\NO{\Gamma_{\alpha,\beta}(z,w) \pd \phi_\gamma(w)}
\bigg)
\epsilon(\alpha,\beta)+ \per
+ \dlt{-(\alpha,\beta)+1}
\big((\alpha,\gamma)(\beta,\rho)-(\rho,\gamma) \big)
\Gamma_{\alpha,\beta}(z,w) \epsilon(\alpha,\beta).
\end{align}
Here $\Gamma_{\alpha,\beta}(z,w)$ stands for
\begin{equation}
                   \label{Gamma_alpha_beta}
\Gamma_{\alpha,\beta}(z,w)=c_{\alpha+\beta}
\NO{\ee{(\phi_\alpha(z)+\phi_\beta(w))}}.
\end{equation}
In (\ref{Gamma-Gamma}) we wrote the result explicitly for
$(\alpha,\beta)\ge 0$
to emphasize the vanishing of the commutator,
in (\ref{Gamma-Lambda}) and (\ref{Lambda-Lambda})
we used a convention $\dlt{n}=0$ if $n<0$ to save the space.

Just to illustrate the calculation of the commutators we want
to prove here the formula~(\ref{Gamma-Gamma}).
First we use the identity~(\ref{exp+exp-}) to show
that
\begin{equation}
\NO{\ee{\phi_\alpha(z)}} \NO{\ee{\phi_\beta(w)}}=
\NO{\ee{(\phi_\alpha(z)+\phi_\beta(w))}}
s_{z>w} (z-w)^{(\alpha,\beta)},
\end{equation}
and then obviously
$
\NO{\ee{\phi_\beta(w)}} \NO{\ee{\phi_\alpha(z)}} =
\NO{\ee{(\phi_\alpha(z)+\phi_\beta(w))}}
s_{w>z} (w-z)^{(\alpha,\beta)}
$.
Using the definitions~(\ref{cocycle_product})
and~(\ref{Gamma_alpha_beta})
one then easily obtains
\begin{equation}
\Gamma_\alpha(z) \Gamma_\beta(w)=
\epsilon(\alpha,\beta)
\Gamma_{\alpha,\beta}(z,w)
s_{z>w} (z-w)^{(\alpha,\beta)},
\end{equation}
and similarly for $\Gamma_\beta(w)\Gamma_\alpha(z)$.
In the case
$(\alpha,\alpha) (\beta,\beta)\in 2\mathbb{Z}$
the formula~(\ref{cocycle_3}) reads
$
\epsilon(\alpha,\beta)=
(-1)^{(\alpha,\beta)}\epsilon(\beta,\alpha).
$
So we finally get
\begin{equation}
\com{\Gamma_\alpha(z)}{ \Gamma_\beta(w)}
=
\epsilon(\alpha,\beta)
\Gamma_{\alpha,\beta}(z,w)
\Big(
s_{z>w} (z-w)^{(\alpha,\beta)}
-s_{w>z} (z-w)^{(\alpha,\beta)}
\Big).
\end{equation}
The expression $s_{z>w} (z-w)^{(\alpha,\beta)}
-s_{w>z} (z-w)^{(\alpha,\beta)}$ vanishes if
$(\alpha,\beta)\ge 0$, and is equal to a derivative of
a delta-function by definition~(\ref{delta_der_def}).
So we have proved the formula~(\ref{Gamma-Gamma}).

Now we want to list more specific commutators which
we need to compute the commutator of the expression
in our ansatz~(\ref{main}). We fix now the root lengths
in the vertex operators $\Gamma_{\alpha}(z)$ ($(\alpha,\alpha)=4$)
and $\Lambda_{\beta,\gamma}(z)$ ($(\beta,\beta)=2$),
and list the mutual commutation relations for different
values of products between different roots.

The first set of commutation relations is essentially the same as
one in~(\ref{Omega1},\ref{Omega2},\ref{Omega3}).
\begin{align}
                      \label{Omega-Omega}
\com{\Omega_A(z)}{\Omega_A(w)}&=\frac{1}{2} \dlt{3}\mathrm{Tr}(A^2)
\per
+\dlt{1}2\,\Omega_{A^2}(w) +\dterm,\\
                      \label{Omega-Gamma}
\com{\Omega(z)}{\Gamma_\alpha(w)}+
\com{\Gamma_{\alpha}(z)}{\Omega(w)}&=
\dlt{1}(\alpha,A\alpha) \Gamma_\alpha(w) +\dterm,\\
                      \label{Omega-Lambda}
\com{\Omega(z)}{\Lambda_{\alpha,\gamma}(w)}+
\com{\Lambda_{\alpha,\gamma}(z)}{\Omega(w)}&=
\dlt{1}\Lambda_{\alpha,y}+\dterm, \text{ where } \per
y=(\alpha,A\alpha)\gamma-(\alpha,A\gamma)\alpha+2 A\gamma .
\end{align}

The vertex operators corresponding to the long roots
($(\alpha,\alpha)=(\beta,\beta)=4$) give rise to the following commutation
relations:
\begin{align}
                       \label{long_long_self}
\com{\Gamma_\alpha(z)}{\Gamma_{-\alpha}(w)}+
\com{\Gamma_{-\alpha}(z)}{\Gamma_\alpha(w)}&=
2\dlt{3}- \per
-\dlt{1} \NO{\pd \phi_\alpha(w) \pd \phi_\alpha(w)}+
\dterm,
\end{align}
\begin{equation}
                     \label{long_long_table}
\begin{array}{|c|c|}
\hline
(\alpha,\beta)&
\vst
\com{\Gamma_\alpha(z)}{\Gamma_\beta(w)}+
\com{\Gamma_\beta(z)}{\Gamma_\alpha(w)}
\\
\hline
-3 &
\vst
\ii \dlt{1} \Lambda_{\alpha+\beta,\alpha-\beta}(w) \epsilon(\alpha,\beta)+
\dterm\\\hline
-2 & \vst
2 \dlt{1} \Gamma_{\alpha+\beta}(w) \epsilon(\alpha,\beta) +\dterm\\
\hline
-1 & \vst 0 \\
\hline
 \ge 0 & \vst 0 \\
\hline
\end{array}
\end{equation}
In the case $(\alpha,\beta)=-3$ the result is different from
the naively expected one: the most singular term is $\dlt{1}$ and not
$\dlt{2}$. This is due to a cancelation of the $\dlt{2}$ terms
coming from
$\com{\Gamma_\alpha(z)}{\Gamma_{\beta}(w)}$ and
$\com{\Gamma_{\beta}(z)}{\Gamma_\alpha(w)}$.
A similar cancelation happens in the $(\alpha,\beta)=-1$ case.

The long root vertex operator $\Gamma_\alpha(z), (\alpha,\alpha)=4$
and the short root vertex operator
$\Lambda_{\beta,\gamma}(z), (\beta,\beta)=2, (\beta,\gamma)=0$
give rise to the following commutation relations:
\begin{equation}
                       \label{long_short_table}
\begin{array}{|c|c|}
\hline
(\alpha,\beta)&
\vst
\com{\Gamma_\alpha(z)}{\Lambda_{\beta,\gamma}(w)}+
\com{\Lambda_{\beta,\gamma}(z)}{\Gamma_\alpha(w)}
\\
\hline
-2 &
\vst
2 \dlt{1} \Lambda_{\alpha+\beta,y}(w) \epsilon(\alpha,\beta)+
\dterm, \text{ where } y=\gamma-\frac{(\alpha,\gamma)}{2}(\alpha-\beta)\\
\hline
-1 & \vst
2 \ii \dlt{1} (\alpha,\gamma) \Gamma_{\alpha+\beta}(w) \epsilon(\alpha,\beta) +\dterm\\
\hline
\ge 0 & \vst 0 \\
\hline
\end{array}
\end{equation}
And the case of two short root vertex operators:
$\Lambda_{\alpha,\rho}(z)$ and $\Lambda_{\beta,\gamma}(z)$,
where $(\alpha,\alpha)=(\beta,\beta)=2,$ $(\alpha,\rho)=(\beta,\gamma)=0$.
\begin{align}
                  \label{short_short_self}
\com{&\Lambda_{\alpha,\rho}(z)}{\Lambda_{-\alpha,\gamma}(w)}+
\com{\Lambda_{-\alpha,\gamma}(z)}{\Lambda_{\alpha,\rho}(w)}=
2(\rho,\gamma)\dlt{3}- \per
-\dlt{1}\big(
2\NO{\pd \phi_\rho(w) \pd \phi_\gamma(w)}+
(\rho,\gamma) \NO{\pd \phi_\alpha(w) \pd \phi_\alpha(w)}
\big)+\dterm,
\end{align}
\begin{equation}
                       \label{short_short_table}
\begin{array}{|c|c|}
\hline
(\alpha,\beta)&
\vst
\com{\ii \Lambda_{\alpha,\rho}(z)}{\ii\Lambda_{\beta,\gamma}(w)}+
\com{\ii\Lambda_{\beta,\gamma}(z)}{\ii\Lambda_{\alpha,\rho}(w)}\\
\hline
-1 &
\begin{array}{l}
\vst -\ii \dlt{1} \Lambda_{\alpha+\beta,y}\epsilon(\alpha,\beta)
+\dterm, \text{ where }\\
y=\big((\alpha,\gamma)(\beta,\rho)-(\rho,\gamma)\big)
(\alpha-\beta)+2(\alpha,\gamma)\rho-2(\beta,\rho)\gamma
\end{array}\\
\hline
0& \vst
-2\dlt{1}\Gamma_{\alpha+\beta}(w)
\big((\alpha,\gamma)(\beta,\rho)-(\rho,\gamma)
\big)
\epsilon(\alpha,\beta)+\dterm\\
\hline
> 0& \vst 0 \\
\hline
\end{array}
\end{equation}

The formulae~(\ref{Omega-Omega}-\ref{short_short_table}) show that
for the field $L(z)$ in ansatz~(\ref{main}) the commutator
$\com{L(z)}{L(w)}$ has the same structure as the Virasoro commutation
relation~(\ref{Virasoro relation}), namely:
\begin{itemize}

\item

There are terms with singularities $\dlt{3}, \dlt{1}, \dl$,
but the term at $\dlt{2}$ cancels out.

\item

The term at $\dlt{3}$ is just a numerical coefficient.

\item

The expression at $\dlt{1}$ consists of terms of the same
type as in $L(z)$ in~(\ref{main}):
$\Omega_B(w), \Gamma_\alpha(z), \Lambda_{\beta,\gamma_\beta}(z)$
for some symmetric matrix $B$, long vectors $\alpha$,
short vectors $\beta$, and  vectors $\gamma_\beta$ such that
$(\gamma_\beta,\beta)=0$.

\end{itemize}

We want to stress again that the term at $\dl$ in the commutator
$\com{L(z)}{L(w)}$ is not an independent one, it is equal
to half the derivative of the term at $\dlt{1}$
for any choice of the field $L(z)$ (of scaling dimension 2).
Hence, one doesn't have to check the term at $\dl$.
So we already see that the expression in our ansatz~(\ref{main})
is a good candidate to satisfy the Virasoro algebra commutation relation.
But the exact conditions on $L(z)$ are still to be specified.
The necessary condition is formulated in the next section.


\section{Consistency conditions for the roots}

\label{Consistency conditions for the roots}

\setcounter{equation}{0}


The procedure of constructing the Virasoro algebra realizations
in the form~(\ref{main})
consists of two parts:
\begin{enumerate}
\item
Find a consistent system of roots $\alpha$ and $\beta$.
\item
Solve the set of quadratic equations for the coefficients.
\end{enumerate}
In this section we focus on the first part of the problem.

It is clear that there are many restrictions on the sets of roots
in~(\ref{main}). The most obvious one is that the products
between the roots should be integer, otherwise the
commutator~(\ref{Virasoro relation}) would not be well defined
due to nonlocality of vertex operators.

There are many other restrictions. For illustration
let us consider a specific case when the expression $L(z)$
in~(\ref{main}) contains two vertex operators
$\Gamma_{\alpha_1}(z)$ and $\Gamma_{\alpha_2}(z)$
and $(\alpha_1,\alpha_1)=(\alpha_2,\alpha_2)=4 $,
$(\alpha_1,\alpha_2)=-2$. Then following the formulae
in~(\ref{long_long_table})
the commutator
$\com{L(z)}{L(w)}$ should include the term
$2 \dlt{1} \Gamma_{\alpha_1+\alpha_2}(w) \epsilon(\alpha_1,\alpha_2)$.
Since the $\dlt{1}$ term in the Virasoro commutator
is just (twice) the Virasoro field itself we deduce that the
field $L(z)$ should also include the vertex operator
$\Gamma_{\alpha_1+\alpha_2}(z)$. Then clearly the long root
$\alpha_1+\alpha_2$ should also belong to the set of roots.

By similar analysis we consider all the possible integer products
between long and short roots. Using the
tables~(\ref{long_long_table}, \ref{long_short_table}, \ref{short_short_table})
we obtain the necessary
conditions for the set of roots to allow the Virasoro algebra realization
in the form~(\ref{main}).
This way we prove the following
\begin{description}
\item[\bf Theorem.] The necessary condition for the field $L(z)$ in (\ref{main})
to satisfy the Virasoro algebra commutation relation is that the set
of roots $\alpha$ and $\beta$ forms a quasi root system.
\end{description}
Here all the consistency conditions for the set of roots are
included in the new notion of a quasi root system,
which we are going to define now.


\section{Definition of quasi root system}

\label{Definition of quasi root system}

\setcounter{equation}{0}


Quasi root system $Q$ is a subset of Euclidean space $\mathbb{E}^n$ with a
standard scalar product $(\ {,}\ )$ provided the following axioms
are satisfied:
\begin{enumerate}

\item
\label{axiom1}

$Q$ is finite, spans $\mathbb{E}^n$, $0 \notin Q$;

\item
\label{axiom2}

If $\alpha \in Q$ then $-\alpha \in Q$, and no other proportional
vectors are in $Q$;

\item
\label{axiom3}

There are two allowed vector lengths: $(\alpha{,}\alpha)=4$ or $2$
(long and short roots);

\item
\label{axiom4}

$(\alpha{,}\beta)\in\mathbb{Z}$ for any $\alpha , \beta \in Q$;


\item
\label{axiom5_rules}

The ``rules'':
\begin{equation}
\begin{array}{|lll|}
\hline
 \alpha_1,\alpha_2 \text{ are long}  &
 { (\alpha_1{,}\alpha_2)=0} & {\raisebox{2pt}{\rule{30pt}{1pt}}}\\
 &{\blue(\alpha_1{,}\alpha_2)=-1} &{\blue \raisebox{2pt}{\rule{30pt}{1pt}}}\\
 &{ (\alpha_1{,}\alpha_2)=-2} &  { \alpha_1 + \alpha_2 \in Q }\\
&{\blue(\alpha_1{,}\alpha_2)=-3} & {\blue\alpha_1 + \alpha_2 \in Q}
\\ \hline
  \alpha  \text{ is long},  \beta  \text{ is short}  &
 {(\alpha{,}\beta)=0 }& {\raisebox{2pt}{\rule{30pt}{1pt}}}\\
 &{\blue(\alpha{,}\beta)=-1} & {\blue\alpha + \beta \in Q} \\
& {(\alpha{,}\beta)=-2} & {\alpha + \beta, \alpha + 2\beta \in Q}
\\ \hline
\beta_1, \beta_2  \text{ are short} &
 {(\beta_1{,}\beta_2)=0} &
{ \beta_1 + \beta_2 \text{ may be in }Q }\\
& {(\beta_1{,}\beta_2)=-1} & {\beta_1 + \beta_2 \in Q}
\\ \hline
\end{array}
\end{equation}

\item
\label{axiom6}

At least one long root is in $Q$.

\end{enumerate}

The axioms \ref{axiom1}-\ref{axiom3} just reflect the setup
of the ansatz~(\ref{main}).
Axiom~\ref{axiom4} follows from the requirement of locality
of the Virasoro field.

Axiom~\ref{axiom5_rules} lists
all the allowed non-positive products between different roots
and the consistency conditions (the ``rules'') for each product.
We stress that all the geometrically allowed non-positive integer
products are in the list.
The ``rules'' are the result of the calculations
in section~\ref{long formulas}
summarized in tables~(\ref{long_long_table}, \ref{long_short_table},
\ref{short_short_table}).
The dash {\raisebox{2pt}{\rule{30pt}{1pt}}} in the table (Axiom~\ref{axiom5_rules})
means that there is no ``rule'' for the given product between
the roots, i.e.~no consistency conditions are applicable.
To be more specific we can tell that the sum of corresponding roots
is not in $Q$ in these cases, since it would be of wrong length.
For the case of two perpendicular short roots also there is no rule,
but in this case the sum of the roots may belong to the quasi root
system, but may be out of it as well.
Remember that all the roots come in pairs of opposite vectors,
so the positive products are of course also allowed,
but we choose to associate the rules with the non-positive products.

Axiom~\ref{axiom6} is needed to exclude the possibility of
 quasi-root systems consisting of short vectors only.
 Multiplying all the vectors by a factor of $\sqrt{2}$
 one would obtain another valid quasi-root system consisting
 of long vectors only. This would contradict the
 philosophy of root systems.
We anticipate that
the ansatz~(\ref{main}) containing $\Lambda$ terms
but no $\Gamma$ terms has no solutions, but
we have checked this fact in the simplest case only.

If one excludes forcedly the following entries
from the list of allowed products between the roots
in Axiom~\ref{axiom5_rules}:
$(\alpha_1,\alpha_2)=-1,-3$ and $(\alpha,\beta)=-1$,
by so modifying Axiom~\ref{axiom5_rules} to:
\begin{equation}
\begin{array}{|lll|}
\hline
 \alpha_1,\alpha_2 \text{ are long}  &
 { (\alpha_1{,}\alpha_2)=0} & {\raisebox{2pt}{\rule{30pt}{1pt}}}\\
 &{ (\alpha_1{,}\alpha_2)=-2} &  { \alpha_1 + \alpha_2 \in Q }
\\ \hline
  \alpha  \text{ is long},  \beta  \text{ is short}  &
 {(\alpha{,}\beta)=0 }& {\raisebox{2pt}{\rule{30pt}{1pt}}}\\
& {(\alpha{,}\beta)=-2} & {\alpha + \beta, \alpha + 2\beta \in Q}
\\ \hline
\beta_1, \beta_2  \text{ are short} &
 {(\beta_1{,}\beta_2)=0} &
 { \beta_1 + \beta_2 \text{ may be in }Q }\\
& {(\beta_1{,}\beta_2)=-1} & {\beta_1 + \beta_2 \in Q}
\\ \hline
\end{array}
\end{equation}
but leaves all the other axioms unmodified,
then one essentially obtains a definition of a Cartan type
root system
except $G_2$ (which is excluded by axiom~\ref{axiom3}),
i.e.~$A_n,\, n \ge 1$,
$B_n,\, n \ge 2$,
$C_n,\, n \ge 3$,
$D_n,\, n \ge 4$,
$F_4, E_6, E_7, E_8$.
In our case there are more allowed values of products between
different roots, so we expect to find additional quasi root systems
different from the Cartan root systems.
Indeed there exist many new quasi root systems as we see in the next section.


\section{Examples of quasi root systems}

\label{Examples of quasi root systems}

\setcounter{equation}{0}


We discuss here only simple quasi root systems,
those that cannot be decomposed to two sets of roots such that
all roots in one set are orthogonal to all roots in the second set.

According to our definition
 a root system of any simple Lie algebra except $G_2$
is a quasi root system. So the
$A_n,\, n \ge 1$,
$B_n,\, n \ge 2$,
$C_n,\, n \ge 3$,
$D_n,\, n \ge 4$,
$F_4, E_6, E_7, E_8$ root systems
are quasi root systems.

Now we want to list low dimensional 
quasi root systems.

\vspace{3pt}\noindent
{$\mathbf{d=1}.$}
There is only one one-dimensional quasi root system: the $A_1$ root system,
a root system
of the $sl_2$ Lie algebra. It consists of two vectors: $\alpha$ and $-\alpha$
with $(\alpha,\alpha)=4$.

\vspace{3pt}\noindent
{$\mathbf{d=2}.$}
There are 4 two-dimensional quasi root systems.
\begin{enumerate}
\addtolength{\parskip}{-5pt}

\item
The $A_2$ root system (figure~\ref{figure_A2}) is generated by two long
roots $(\alpha_1,\alpha_2)=-2$, the third root $\alpha_3=\alpha_1+\alpha_2$
is also a long one. There are 6 vectors in the root system. It is a root system
of the $sl(3)$ Lie algebra.
\begin{figure}[h]
\begin{center}
\includegraphics[width=100pt]{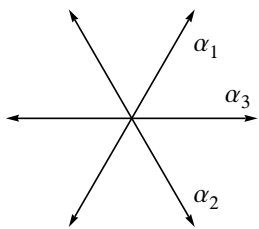}
\end{center}
\caption{$A_2$ root system: $(\alpha_1,\alpha_2)=-2$, $(\alpha_i,\alpha_i)=4$.}
\label{figure_A2}
\end{figure}

\item
The $B_2$ root system (figure~\ref{figure_B2}) consists of 8 roots --
4 long and 4 short. The generating roots are $\alpha_2$ (long)
and $\beta_2$ (short), such that $(\alpha_2,\beta_2)=-2$.
Other roots are obtained as $\beta_1=\alpha_2+\beta_2$,
$\alpha_1=\alpha_2+2\beta_2$. $B_2$ is a root system of the $so(5)$ Lie algebra.

\begin{figure}[h]
\begin{center}
\includegraphics[width=100pt]{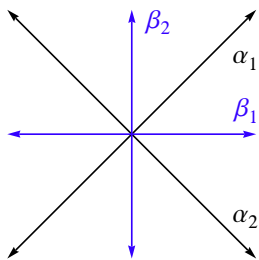}
\end{center}
\caption{$B_2$ root system: $(\alpha_2,\beta_2)=-2$, $(\alpha_2,\alpha_2)=4$, $(\beta_2,\beta_2)=2$.}
\label{figure_B2}
\end{figure}

\item
The third two-dimensional quasi root system
(figure~\ref{figure_T2}), which we call   $T_2$, consists of 6 roots --
4 long, 2 short. The product between two long roots is
$(\alpha_1,\alpha_2)=-3$. The short root is
$\beta=\alpha_1+\alpha_2$.
\begin{figure}[h]
\begin{center}
\includegraphics[width=100pt]{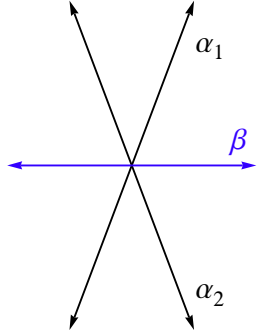}
\end{center}
\caption{$T_2$ quasi root system: $(\alpha_1,\alpha_2)=-3$, $(\alpha_i,\alpha_i)=4$.}
\label{figure_T2}
\end{figure}

\item
The last two-dimensional quasi root system
(figure~\ref{figure_I2}) is probably the simplest one.
It contains just two pairs of opposite roots, 4 long roots altogether.
The product is
$(\alpha_1,\alpha_2)=-1$. We call this system the $I_2$ quasi root system.
\begin{figure}[h]
\begin{center}
\includegraphics[width=100pt]{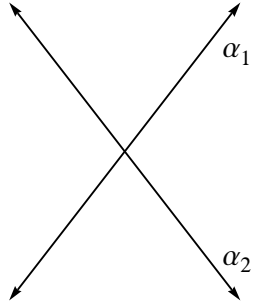}
\end{center}
\caption{$I_2$ quasi root system: $(\alpha_1,\alpha_2)=-1$, $(\alpha_i,\alpha_i)=4$.}
\label{figure_I2}
\end{figure}

\end{enumerate}


%

%

\vspace{3pt}\noindent
{$\mathbf{d=3}.$}
There are 15 three-dimensional quasi root systems.
We list them in Appendix~\ref{QR_dim3}.

\vspace{3pt}\noindent
{$\mathbf{d \ge 4}.$}
We iteratively generated quasi root systems on a computer:
there are 83 four-dimensional quasi root systems,
at least 534 five-dimensional,
a few thousands of 6-dimensional quasi root systems.
The number of quasi root systems grows very fast with
dimension.

The next set of examples is given by quasi root systems
formed by 
long roots only with all the products between
different roots being 0, $-1$ or 1.
Since the ``rules'' in Axiom~\ref{axiom5_rules} in the definition
of a quasi root system imply no consistency conditions on pairs
of long roots with product 0 or $\pm 1$ such set of long roots
does constitute a quasi root system.
The matrix of scalar products between the roots
is a symmetric nonnegative definite matrix
with all the diagonal elements equal to 4,
and all the off-diagonal elements equal to 0 or $\pm 1$.
We will refer to these systems as quasi root systems of type $I$.

Every symmetric positive definite $n\times n$ dimensional matrix
with 4 on the diagonal and 0 or $\pm 1$ off diagonal gives rise
to a quasi root system of $i$ type formed by $n$ pairs of long roots.
We also want to present an explicit example when the matrix
is degenerate.
First take a four dimensional quasi root
system generated by $\alpha_i,\ i=1,2,3,4$:
\begin{equation}
(\alpha_i,\alpha_j)=\left\{
\begin{array}{ll}
4, & i=j,\\
-1, & i \ne j.
\end{array}
\right.
\qquad
i,j=1,2,3,4.
\end{equation}
This quasi root system contains 8 roots (remember the opposite roots!).
One obtains another valid four dimensional quasi root system
by fitting one more pair of long opposite roots:
\begin{equation}
\alpha_5=\alpha_1 +\alpha_2 +\alpha_3 +\alpha_4.
\end{equation}
The root $\alpha_5$ is long, and the product $(\alpha_5,\alpha_i)=1$,
$i=1,2,3,4$. There is a total of 10 roots in the new quasi root system.
The matrix of scalar product for the extended system
\begin{equation}
\left(
\begin{array}{ccccc}
4& -1&-1&-1 &1\\
-1&4&-1&-1 &1 \\
-1&-1&4& -1 &1\\
-1&-1&-1 &4& 1\\
1&1&1&1&4
\end{array}
\right)
\end{equation}
is of course degenerate.

Another set of examples of quasi root systems
we borrow from~\cite{Dunne:1988ii}.
Let $\Delta_1$ and $\Delta_2$ be the root systems of $g_1$ and $g_2$,
two simple Lie algebras of ADE type. The roots
$\alpha_i^{(1)}, \alpha_j^{(2)}$ are
normalized to $(\alpha_i^{(1)}, \alpha_i^{(1)})= ( \alpha_j^{(2)},\alpha_j^{(2)})=2 $.
Let $r_1, r_2$ be the dimensions of the root systems (= the ranks of the algebras),
$n_1$ and $n_2$ be the number of roots in $\Delta_1$ and $\Delta_2$
respectively.
Consider the set $Q=\Delta_1\times \Delta_2$ formed by
all the pairs of roots $\alpha_{i j}=\{\alpha_i^{(1)}, \alpha_j^{(2)}\}$
modulo identification
$\{\alpha_i^{(1)}, \alpha_j^{(2)}\}=\{-\alpha_i^{(1)}, -\alpha_j^{(2)}\}$.
The scalar product is defined as
$\big(\{\alpha^{(1)}, \beta^{(2)}\},\{\gamma^{(1)}, \delta^{(2)}\}\big)=
(\alpha^{(1)},\gamma^{(1)})(\beta^{(2)},\delta^{(2)})$.
It is easy to see that Q is a quasi root system of dimension $r_1\, r_2$,
it is formed by $n_1 n_2/2$ long roots with scalar products between
them $0,\pm 1,\pm 2$.

The last example which we list here is a ``bad'' example.
This quasi root system was suggested by Anthony Joseph.
Take 7 mutually orthogonal copies of the following quasi root system:
$\alpha,\beta,\gamma$ are long, $(\alpha,\beta)=(\beta,\gamma)=1$, $(\alpha,\gamma)=0$.
Add another root $\delta$ defined by $14\delta=\sum_{i=1}^7 (3\alpha_i+2\beta_i+3\gamma_i)$.
Then $(\delta,\delta)=4$, $(\delta,\alpha_i)=(\delta,\beta_i)=(\delta,\gamma_i)=1$.
This way one obtains a 21-dimensional quasi root system $Q$ which consists of
$6\cdot 7+2=44$ long roots.
This quasi root system has no integral basis:
there is no basis $\Omega \subset Q$ such that
$\alpha \in \mathbb{Z} \Omega$ for all $\alpha \in Q$.
So we see that in contrast to the root systems of simple Lie algebras
quasi root systems in general do not
admit an integer basis.


\section{Solutions for the ansatz}

\label{Solutions for the ansatz}

\setcounter{equation}{0}


Now we get to the second part of the problem: given the set of roots
$\alpha$ and $\beta$ in the ansatz~(\ref{main}) constitute a quasi
root system one still have to adjust the coefficients
in the expression to satisfy the Virasoro algebra.
We are going to discuss the equations for the coefficients
and to solve these equations for a number of quasi root systems.


\subsection{General discussion}

\label{Solutions for the ansatz: general discussion}


Here we perform a general analysis of quadratic equations
for the coefficients in the ansatz~(\ref{main}).
One substitutes the expression to the Virasoro algebra
commutation relation and obtains a set of equations.
Some of them one can write without specifying the quasi root system.
The equations for the $\NO{\pd \phi \pd \phi}$ type terms
can be easily derived (use formulae~(\ref{Omega-Omega}, \ref{long_long_self},
\ref{short_short_self})):
\begin{equation}
                    \label{phi_phi_equation}
A_{\mu \nu}=(A^2)_{\mu \nu}+\sum_{\alpha>0}b_\alpha^2 \alpha_\mu \alpha_\nu+
2\sum_{\beta>0}(\gamma_\beta)_\mu (\gamma_\beta)_\nu+
\sum_{\beta>0} (\gamma_\beta,\gamma_\beta) \beta_\mu \beta_\nu.
\end{equation}
Here $\alpha_\mu$ stands for the $\mu$ component of the vector $\alpha$.
The summations are only on one root from each pair of opposite
roots, we denote this schematically by ${\alpha>0}, {\beta>0}$
 under the summation symbols. We get $d(d+1)/2$ independent equations,
equal to the number of independent elements $A_{\mu \nu}$ in the symmetric
matrix $A$.
The same formulae~(\ref{Omega-Omega}, \ref{long_long_self},
\ref{short_short_self})
give the expression for the central charge of the Virasoro
algebra:
\begin{equation}
c=\mathrm{Tr}(A^2)+4\sum_{\beta>0} b_\alpha^2+4\sum_{\beta>0} (\gamma_\beta,\gamma_\beta).
\end{equation}
Taking the trace of the matrix equation~(\ref{phi_phi_equation})
we obtain the following simple formula for the central charge:
\begin{equation}
                  \label{central_charge_Tr}
c=\mathrm{Tr}(A).
\end{equation}

Now let us discuss the equations arising from $\Gamma_\alpha$ terms
in the Virasoro algebra commutation relation.
Use formula~(\ref{Omega-Gamma}) to get
\begin{equation}
                      \label{Gamma_equation}
2b_\alpha=(\alpha,A \alpha) b_\alpha
+\text{ inter-vertex contributions. }
\end{equation}
The ``inter-vertex contributions'' are terms coming from the commutators
between the fields of $\Gamma$ and $\Lambda$ type. One has to go over all pairs
of roots which sum to the root $\alpha$ in the equation~(\ref{Gamma_equation}).
The ``inter-vertex contributions'' are obtained then
from the tables~(\ref{long_long_table}, \ref{long_short_table}, \ref{short_short_table}).
There is one equation for each $b_\alpha$ coefficient.

A similar equation for the $\Lambda_{\beta,\gamma_\beta}$ terms reads
\begin{equation}
                      \label{Lambda_equation}
2 \, \ii \gamma_\beta=\ii \big( (\beta,A\beta)\gamma_\beta-
(\beta,A\gamma_\beta)\beta+2A\gamma_\beta \big)
+\text{ inter-vertex contributions. }
\end{equation}
This is a vector equation. The number of independent components
in the vector $\gamma_\beta$ is $d-1$ since it is perpendicular
to $\beta$. But all the terms in the vector equation above
including the ``inter-vertex contributions'' are also perpendicular
to the root $\beta$. So there are $d-1$ equations for $d-1$ variables
(for each $\Lambda$ term).

We conclude that generically we have the same number of equations
and unknown variables.

Now we want to discuss the ``duality'' of the solutions.
The solutions for the ansatz always come in pairs of commutative
Virasoro fields. Suppose that the field $L^{(1)}(z)$ in the form
(\ref{main}) satisfies the Virasoro algebra relations with central
charge $c^{(1)}$,
and the field $L^\text{st}(z)$ is the canonical free boson
realization~(\ref{canonical construction})
of the Virasoro algebra (of central charge $c=d$).
Then the field
\begin{equation}
L^{(2)}(z)=L^\text{st}(z)-L^{(1)}(z)
\end{equation}
also satisfies the Virasoro algebra relations with central charge
$c^{(2)}=d-c^{(1)}$, and the two solutions are commutative:
\begin{equation}
\com{L^{(1)}(z)}{L^{(2)}(w)}=0.
\end{equation}
The proof is simple. First note that
\begin{equation}
\com{L^\text{st}(z)}{L^{(1)}(w)}=\frac{\mathrm{Tr}(A^{(1)})}{2}\dlt{3}+2L^{(1)}(w)\dlt{1}+\pd L^{(1)}(w)\dl.
\end{equation}
Take into account that due to
the central charge formula~(\ref{central_charge_Tr})
we have $\mathrm{Tr}(A^{(1)})=c^{(1)}$.
Now it is easy to calculate $\com{L^{(2)}(z)}{L^{(2)}(w)}$
and $\com{L^{(1)}(z)}{L^{(2)}(w)}$ to verify the statement.

The ``dual'' solution is obtained by flipping the signs of
$b_\alpha$, $\gamma_\beta$, and off-diagonal $A_{\mu \nu}$,
while the diagonal elements are obtained as
$A_{\mu \mu}^{(2)}=1-A_{\mu \mu}^{(1)}$.


\subsection{Solutions for 2-dimensional quasi root systems}


We consider first the low-dimensional examples.
The one-dimensional example (which is based on the $A_1$ root system)
was discussed in
Section~\ref{Vertex operator realization of the Virasoro algebra},
where we saw that the construction reproduces the Ising model ($c=1/2$).
Here we analyze the four two-dimensional quasi root systems.
The interesting cases come from $A_2$ and $I_2$ quasi root systems.
We get the solutions with central charges $c=7/10,4/5,6/5,13/10$
for $A_2$ root system and $c=6/7,8/7$ for $I_2$ quasi root system.
The other two quasi root systems, $B_2$ and $T_2$, both involve
short roots, and so the ansatz expression~(\ref{main}) includes
$\Lambda$ terms. The central
charges for the solutions are $c=1/2,3/2$ in both cases.
For higher dimensional quasi root systems we expect to obtain
more interesting solutions for systems involving short roots, since
there will be more ``space'' to put the $\gamma$ vectors in a nontrivial way.


\subsubsection{$A_2$ root system}

\label{$A_2$ root system}


The $A_2$ root system (Figure~\ref{figure_A2})
is generated by two roots: $\alpha_1, \alpha_2$.
In our normalization the mutual products between
the roots are $(\alpha_1,\alpha_1)=(\alpha_2,\alpha_2)=4$,
$(\alpha_1,\alpha_2)=-2$. The third root is
$\alpha_3=\alpha_1+\alpha_2$. Taking into account the opposite roots
we get 6 roots in the $A_2$ root system.

Since all the mutual products between the roots are even integers
the two-cocycle is trivial:
\begin{equation}
\epsilon(\pm \alpha_i, \pm \alpha_j)=1, \quad i,j=1,2,3.
\end{equation}

The ansatz~(\ref{main}) for the $A_2$ root system reads
\begin{equation}
L=\Omega_A
+b_1\left(\Gamma_{\alpha_1}+\Gamma_{-\alpha_1}\right)
+b_2\left(\Gamma_{\alpha_2}+\Gamma_{-\alpha_2}\right)
+b_3\left(\Gamma_{\alpha_3}+\Gamma_{-\alpha_3}\right),
\end{equation}
where $\Omega_A=
-\frac{1}{2}\displaystyle \sum_{\mu,\nu=\, x\, \text{or}\, y} A_{\mu \nu} \NO{\pd \phi_\mu\pd \phi_\nu}$
We choose the directions $x$ and $y$ along the symmetry axes of the
root system: $x$ in the direction of $\alpha_3$,
then the matrix $A$ can be taken diagonal:
\begin{equation}
\Omega_A=-\frac{1}{2}a_x \NO{\pd \phi_x\pd \phi_x}-\frac{1}{2}
a_y \NO{\pd \phi_y \pd \phi_y}.
\end{equation}

The equations arising from the comparison of $\NO{\pd \phi \pd \phi}$
terms in the Virasoro commutation relation are
\begin{align}
               \label{A_2_1}
0&=b_1^2-b_2^2, \\
               \label{A_2_2}
a_x&=a_x^2+b_1^2+b_2^2+4b_3^2,\\
               \label{A_2_3}
a_y&=a_y^2+3b_1^2+3b_2^2.
\end{align}
And the equations coming from the comparison of $\Gamma_{\alpha_i}$
terms are
\begin{align}
               \label{A_2_4}
2b_3&=4a_x b_3+2b_1 b_2, \\
               \label{A_2_5}
2b_2&=(a_x+3a_y)b_2+2 b_1 b_3,\\
               \label{A_2_6}
2b_1&=(a_x+3a_y)b_1+2 b_2 b_3.
\end{align}
From~(\ref{A_2_1}) one obtains that $b_1=\pm b_2$.
Note that the equations~(\ref{A_2_1}-\ref{A_2_6})
are invariant under the change $b_2 \to -b_2, b_3\to -b_3$.
So we may assume that $b_1=b_2$, and later get all the solutions
by reverting the signs of $b_2$ and $b_3$.

If $b_1=b_2=0$ then the ansatz is reduced to that on $A_1$ root system
plus one free boson. So we will
also assume that $b_1$ and $b_2$ are different from zero.
Then equations~(\ref{A_2_4}) and~(\ref{A_2_5}) give
\begin{align}
                  \label{A_2_b3}
b_3&=1-\frac{1}{2}a_x-\frac{3}{2}a_y,\\
                  \label{A_2_b1}
b_1^2&=b_3(1-2a_x).
\end{align}
Substituting these relations in~(\ref{A_2_2}) and~(\ref{A_2_3})
we obtain:
\begin{align}
\left(2 a_x+3 a_y-3\right)
   \left(2 a_x+3 a_y-2\right)&=0,\\
6 a_x^2+18 a_y a_x-15
   a_x+a_y^2-10 a_y+6&=0.
\end{align}
This set of equations has four solutions:
\begin{equation}
\begin{array}{lll}
a_x = 2/5, & a_y=2/5, & c=4/5;\\
a_x = 3/5, & a_y=3/5, & c=6/5;\\
a_x = 1/10, & a_y=3/5, & c=7/10;\\
a_x = 9/10, & a_y=2/5, & c=13/10;
\end{array}
\end{equation}
where we have included the central charge $c=a_x+a_y$ in each case.
The values of $b_i$ are then calculated using~(\ref{A_2_b3})
and~(\ref{A_2_b1}).
The full set of nontrivial solutions is
\begin{align}
c=4/5: & \quad a_x = 2/5,\ a_y=2/5,\
\begin{array}{lll}
b_1 & b_2 & b_3\\
1/5 & 1/5 & 1/5\\
1/5 & -1/5 & -1/5\\
-1/5 & -1/5 & 1/5\\
-1/5 & 1/5 & -1/5\\
\end{array}
\\
c=6/5:& \quad a_x = 3/5, \ a_y=3/5, \
\begin{array}{lll}
b_1 & b_2 & b_3\\
1/5 & 1/5 & -1/5\\
1/5 & -1/5 & 1/5\\
-1/5 & 1/5 & 1/5\\
-1/5 & -1/5 & -1/5\\
\end{array}
\\
c=7/10:& \quad a_x = 1/10,\ a_y=3/5,\
\begin{array}{lll}
b_1 & b_2 & b_3\\
1/5 & 1/5 & 1/20\\
1/5 & -1/5 & -1/20\\
-1/5 & -1/5 & 1/20\\
-1/5 & 1/5 & -1/20\\
\end{array}
\\
c=13/10:& \quad a_x = 9/10,\ a_y=2/5,\
\begin{array}{lll}
b_1 & b_2 & b_3\\
1/5 & 1/5 & -1/20\\
1/5 & -1/5 & 1/20\\
-1/5 & 1/5 & 1/20\\
-1/5 & -1/5 & -1/20\\
\end{array}
\end{align}


\subsubsection{$I_2$ quasi root system}

\label{$I_2$ quasi root system}


Here we discuss the solutions for the $I_2$ quasi root system, which is shown on
Figure~\ref{figure_I2}. The quasi root system is formed
by two long roots $\alpha_1$ and $\alpha_2$ with product
$(\alpha_1,\alpha_2)=-1$. The 2-cocycle is
$\epsilon(\alpha_1,\alpha_2)=1=-\epsilon(\alpha_2,\alpha_1)$.

The ansatz for the Virasoro field is
\begin{equation}
                   \label{L in -1}
L=\Omega_A
+b_1\left(\Gamma_{\alpha_1}+\Gamma_{-\alpha_1}\right)
+b_2\left(\Gamma_{\alpha_2}+\Gamma_{-\alpha_2}\right).
\end{equation}
We choose again the axes $x$ and $y$ along the symmetry axes
of the root system: $x$ in the direction of $\alpha_1+\alpha_2$,
$y$ in the direction of $\alpha_1-\alpha_2$.
Then the matrix $A$ becomes diagonal:
\begin{equation}
\Omega_A=-\frac{1}{2}a_x \NO{\pd \phi_x\pd \phi_x}-\frac{1}{2}
a_y \NO{\pd \phi_y \pd \phi_y}.
\end{equation}

We substitute the expression~(\ref{L in -1}) to the Virasoro algebra
commutation relation and get the following equations:
from the comparison of $\NO{\pd \phi\pd \phi}$ terms
\begin{align}
0&=b_1^2-b_2^2,\\
a_x&=a_x^2+\frac{3}{2}b_1^2+\frac{3}{2}b_2^2,\\
a_y&=a_y^2+\frac{5}{2}b_1^2+\frac{5}{2}b_2^2,
\end{align}
and from the comparison of $\Gamma$ terms:
\begin{align}
2b_i&=(\frac{3}{2}a_x+\frac{5}{2}a_y)b_i,
\qquad i=1,2.
\end{align}
In the last equation there are no terms that mix different
vertex operators since the relevant commutator
in~(\ref{long_long_table}) is zero.

Assuming that $b_1 \ne 0$ and $b_2 \ne 0$ we reduce the equations
above to one quadratic equation. The solutions are
\begin{align}
c=6/7: & \quad
a_x=1/7,\ a_y=5/7, \\
                \label{I2_c8_7}
c=8/7: & \quad
a_x=6/7,\ a_y=2/7,
\end{align}
and $b_1^2=b_2^2=2/49$ in both cases.


\subsubsection{$T_2$ quasi root system}

\label{$T_2$ quasi root system}


The $T_2$ quasi root system (Figure~\ref{figure_T2})
is generated by two long roots $\alpha_1, \alpha_2$:
$(\alpha_1,\alpha_1)=(\alpha_2,\alpha_2)=4$,
$(\alpha_1,\alpha_2)=-3$. The third root $\beta=\alpha_1+\alpha_2$
is short: $(\beta,\beta)=2, (\beta,\alpha_i)=1$.
There are 6 roots in the root system.

We choose the 2-cocycle as described in Section~\ref{Vertex operators}:
$\epsilon(\alpha_1,\alpha_2)=1=-\epsilon(\alpha_2,\alpha_1)$.
Then $\epsilon(\alpha_1,\beta)=\epsilon(\alpha_1,\alpha_1)\epsilon(\alpha_1,\alpha_2)=1$
and $\epsilon(\alpha_2,\beta)=\epsilon(\alpha_2,\alpha_1)\epsilon(\alpha_2,\alpha_2)=-1$.

The expression for the Virasoro field based on the quasi root system
is
\begin{equation}
L=\Omega_A
+b_1\left(\Gamma_{\alpha_1}+\Gamma_{-\alpha_1}\right)
+b_2\left(\Gamma_{\alpha_2}+\Gamma_{-\alpha_2}\right)
+\ii \left(\Lambda_{\beta,\gamma}-\Lambda_{-\beta,\gamma}\right),
\end{equation}
where $(\gamma,\beta)=0$.

Again we choose the directions $x$ and $y$ along the symmetry axes of the
root system: $x$ in the direction of $\beta$,
$y$ in the direction of $\alpha_1-\alpha_2$.
Then the matrix $A$ can be taken diagonal, so
\begin{equation}
\Omega_A=-\frac{1}{2}a_x \NO{\pd \phi_x\pd \phi_x}-\frac{1}{2}
a_y \NO{\pd \phi_y \pd \phi_y}.
\end{equation}
The $\Lambda$ term can be written as
$\Lambda_{\beta,\gamma}=
g\, \pd \phi_y \Gamma_{\beta}=
g\, c_\beta \, \pd \phi_y  \NO{\ee{\sqrt{2}\phi_x}}$,
where the vector $\gamma$ was substituted by $\gamma=(0,g)$
in coordinates $x,y$.

The comparison of $\NO{\pd \phi \pd \phi}$
terms in the Virasoro commutation relation gives
\begin{align}
               \label{qr3_1}
0&=b_1^2-b_2^2, \\
a_x&=a_x^2+\frac{1}{2}b_1^2+\frac{1}{2}b_2^2+2g^2,\\
a_y&=a_y^2+\frac{7}{2}b_1^2+\frac{7}{2}b_2^2+2g^2.
\end{align}
Equations coming from the comparison of $\Gamma_{\alpha_i}$
terms are
\begin{align}
2b_2&=(\frac{1}{2}a_x+\frac{7}{2}a_y)b_2+\sqrt{14}b_1 g,\\
2b_1&=(\frac{1}{2}a_x+\frac{7}{2}a_y)b_1+\sqrt{14}b_2 g.
\end{align}
And the $\Lambda$ term gives the following equation:
\begin{align}
               \label{qr3_2}
2g&=2(a_x+a_y)g+\sqrt{14}b_1 b_2.
\end{align}
Following the same guidelines as in Section~\ref{$A_2$ root system}
one can easily obtain the nontrivial ($b_1, b_2,g \neq 0$)
solutions to the set of equations~(\ref{qr3_1}-\ref{qr3_2}):
\begin{align}
c=1/2: & \quad a_x = 1/16,\ a_y=7/16,\
\begin{array}{lll}
b_1 & b_2 & g \\
1/(4\sqrt{2}) & 1/(4\sqrt{2}) & \sqrt{14}/32\\
-1/(4\sqrt{2}) & -1/(4\sqrt{2}) & \sqrt{14}/32\\
1/(4\sqrt{2}) & -1/(4\sqrt{2}) & -\sqrt{14}/32\\
-1/(4\sqrt{2}) & 1/(4\sqrt{2}) & -\sqrt{14}/32\\
\end{array}
\\
c=3/2: & \quad a_x = 15/16,\ a_y=9/16,\
\begin{array}{lll}
b_1 & b_2 & g \\
-1/(4\sqrt{2}) & -1/(4\sqrt{2}) & -\sqrt{14}/32\\
1/(4\sqrt{2}) & 1/(4\sqrt{2}) & -\sqrt{14}/32\\
1/(4\sqrt{2}) & -1/(4\sqrt{2}) & \sqrt{14}/32\\
-1/(4\sqrt{2}) & 1/(4\sqrt{2}) & \sqrt{14}/32\\
\end{array}
\end{align}


\subsubsection{$B_2$ root system}


This is maybe the most complex example between all two-dimensional
examples. The $B_2$ root system consists of two pairs of long roots
and two pairs of short roots, see Figure~\ref{figure_B2}.
The 2-cocycle is nontrivial. We choose $\epsilon(\beta_1,\beta_2)=-1$
and $\epsilon(\beta_1,\beta_1)=\epsilon(\beta_2,\beta_2)=-1$,
then $\epsilon(\alpha_i,\beta_j)=1$ and $\epsilon(\alpha_i,\alpha_j)=1$
for $i,j=1,2$.

As usual we choose the coordinate axes along the symmetry axes
of the root system: $x$ along $\beta_1$, $y$ along $\beta_2$.
And we also let the matrix $A$ to be diagonal. So we are looking
for the Virasoro field in the following form:
\begin{equation}
\begin{aligned}
L&=-\frac{1}{2}a_x \NO{\pd \phi_x\pd \phi_x}-\frac{1}{2}
a_y \NO{\pd \phi_y \pd \phi_y}+\\
&+b_1\left(\Gamma_{\alpha_1}+\Gamma_{-\alpha_1}\right)
+b_2\left(\Gamma_{\alpha_2}+\Gamma_{-\alpha_2}\right)+\\
&+\ii g_1 \left(\Lambda_{\beta_1,i_y}-\Lambda_{-\beta_1,i_y}\right)
+\ii g_2 \left(\Lambda_{\beta_2,i_x}-\Lambda_{-\beta_2,i_x}\right),
\end{aligned}
\end{equation}
where $i_x=(1,0)$ and $i_y=(0,1)$ are the identity vectors
in the directions of axes $x$ and $y$.

The equations for the $\NO{\pd \phi \pd \phi}$ terms read
\begin{align}
0&=b_1^2-b_2^2 
,
\\
a_x&=a_x^2 
+2b_1^2+2b_2^2+2g_1^2+2g_2^2,\\
a_y&=a_y^2 
+2b_1^2+2b_2^2+2g_1^2+2g_2^2.
\end{align}
The equations coming from the $\Gamma$ terms:
\begin{align}
b_1\left(
a_x+a_y-1
\right)
+2g_1 g_2&=0
,\\
b_2\left(
a_x+a_y-1
\right)
+2g_1 g_2&=0
.
\end{align}
And the $\Lambda$ terms give the following equations:
\begin{align}
g_1\left(a_x+a_y-1\right)+2g_2\left(b_1+b_2\right)&=0,\\
g_2\left(a_x+a_y-1\right)+2g_1\left(b_1+b_2\right)&=0.
\end{align}

We are looking for solutions with $b_1, b_2, g_1, g_2$
all being different from zero. It is easy to see that then
$b_1=b_2,\ g_1=\pm g_2,\ a_x=a_y$.
And the full set of solutions is
\begin{align}
c=1/2: & \quad a_x = a_y=1/4,\
\begin{array}{ccc}
b_1 = b_2 & g_1 & g_2 \\
1/8 & 1/(4\sqrt{2}) & 1/(4\sqrt{2})\\
1/8 & -1/(4\sqrt{2}) & -1/(4\sqrt{2})\\
-1/8 & -1/(4\sqrt{2}) & 1/(4\sqrt{2})\\
-1/8 & 1/(4\sqrt{2}) & -1/(4\sqrt{2})\\
\end{array}
\\
c=3/2: & \quad a_x = a_y=3/4,\
\begin{array}{ccc}
b_1 = b_2 & g_1 & g_2 \\
1/8 & 1/(4\sqrt{2}) & -1/(4\sqrt{2})\\
1/8 & -1/(4\sqrt{2}) & 1/(4\sqrt{2})\\
-1/8 & 1/(4\sqrt{2}) & 1/(4\sqrt{2})\\
-1/8 & -1/(4\sqrt{2}) & -1/(4\sqrt{2})\\
\end{array}
\end{align}


\subsection{Solutions for $I$-type quasi root systems}

\label{Solutions for $I$-type quasi root systems}


In this section we analyze $d$-dimensional
quasi root systems formed by $d$ pairs of
long opposite roots with all the products between
different roots being 0, $-1$ or 1. In other words
the matrix of scalar products between the $d$ independent roots
is a symmetric positive definite matrix
with all the diagonal elements equal to 4,
and all the off-diagonal elements equal to 0 or $\pm 1$:
\begin{equation}
C_{i j}=(\alpha_i,\alpha_j)=\left\{
\begin{array}{ll}
4, & i=j,\\
-1, & i \ne j.
\end{array}
\right.
\qquad
i,j=1,\ldots,d.
\end{equation}
These quasi root systems are easy to solve since the equations
for the coefficients~(\ref{phi_phi_equation}, \ref{Gamma_equation})
have no ``inter-vertex contributions''.

The equations we want to solve are
\begin{align}
                        \label{pm1_eq1}
A_{\mu \nu}&=(A^2)_{\mu \nu}+\sum_{\alpha>0}b_\alpha^2 \alpha_\mu \alpha_\nu,
\qquad \mu,\nu=1,\ldots,d,\\
                        \label{pm1_eq2}
2b_\alpha&=(\alpha,A \alpha) b_\alpha,
\end{align}
here $\sum_{\alpha>0}$ means that the summation is taken on
``positive'' roots only, i.e.~one root only from each pair
of opposite roots is taken into account.
We are interested in solutions where all $b_\alpha$ are different from zero.
Therefore we can drop $b_\alpha$ in the last equation.

We introduce a $d\times d$ matrix $V$ the columns of which are just
our $d$ independent roots. Then the matrix of scalar products can be expressed as
$C=V^\mathrm{T} V$.
According to~(\ref{pm1_eq2})
the symmetric matrix
\begin{equation}
                \label{pm1_S_A}
S=V^\mathrm{T}\! A V
\end{equation}
has all diagonal elements equal to $2$.
Now we multiply the equation~(\ref{pm1_eq1}) from left by $V^\mathrm{T}$
and from right by $V$, the result is
\begin{equation}
S=S C^{-1} S+C B C,
\end{equation}
where $B$ is a diagonal matrix having $b_{\alpha_1}^2,\ldots,b_{\alpha_d}^2$
on the diagonal. Multiplying by $C^{-1}$ from left and right we get
\begin{equation}
                         \label{CSCSC}
C^{-1}SC^{-1}=C^{-1}S C^{-1} SC^{-1}+ B ,
\end{equation}
Now we have here $d(d+1)/2$ equations for the same number of unknowns:
$d$ unknowns $b_{\alpha_i}^2$ and $d(d-1)/2$ off-diagonal elements of matrix $S$.
(Remember that the diagonal elements of $S$ are known -- all equal to 2.)
Note that we got rid of the coordinate dependence -- $A$ and $V$ do not enter
to the last equation. Considering only off-diagonal terms of the matrix
equation we obtain $d(d-1)/2$ quadratic equations for
off-diagonal elements of matrix $S$. One can solve the set of equations
on a computer (at least numerically). Then $b_{\alpha_i}^2$ are obtained
from the diagonal terms of the same matrix equation. The central charge
is given by $c=\mathrm{Tr}A=\mathrm{Tr}(S C^{-1})$. Then one can calculate
also the matrix $A$ in a given system of coordinates: one has to write down
the matrix $V$ of roots in the chosen system of coordinates and then to use
equation~(\ref{pm1_S_A}) to obtain $A$ from $S$.

The two dimensional example is already discussed in section~\ref{$I_2$ quasi root system},
the matrix $C$ in this case is
\begin{equation}
C=
\left(
\begin{array}{cc}
4 & -1 \\
-1 & 4
\end{array}
\right).
\end{equation}
It is an easy exercise to check that the matrix equation~(\ref{CSCSC})
leads to the same result.

In three dimensions there are 3 simple quasi root systems of $I$ type.
These are listed in Appendix~\ref{QR_dim3} under numbers 4,5,6.
We present here the solutions for all the three systems.
We list the central charge and $b_\alpha^2$ only, the form of matrix
$A$ depends on the choice of coordinate system.

\begin{itemize}

\item

Three dimensional quasi root system number~5 in Appendix~\ref{QR_dim3}.
The $C$ matrix is
\begin{equation}
                     \label{C_forItype_dim3_minus1}
C=
\left(
\begin{array}{ccc}
4 & -1 & -1\\
-1 & 4& -1\\
-1& -1& 4
\end{array}
\right).
\end{equation}
The solutions are
\begin{equation}
                        \label{solution_QR5}
\begin{array}{rlll}
c=\frac{14}{11}: &
b_{\alpha_1}^2 =\frac{4}{121}, & b_{\alpha_2}^2=\frac{6}{121},
& b_{\alpha_3}^2=\frac{6}{121}, \\[5pt]
c=\frac{19}{11}: &
b_{\alpha_1}^2 =\frac{4}{121}, & b_{\alpha_2}^2=\frac{6}{121},
& b_{\alpha_3}^2=\frac{6}{121}, \\[5pt]
c=\frac{3}{2}
\left(
1+\frac{1}{\sqrt{41}}
\right): &
b_{\alpha_1}^2 =\frac{2}{41}, & b_{\alpha_2}^2=\frac{2}{41}, & b_{\alpha_3}^2=\frac{2}{41}, \\[5pt]
c=\frac{3}{2}
\left(
1-\frac{1}{\sqrt{41}}
\right): &
b_{\alpha_1}^2 =\frac{2}{41}, & b_{\alpha_2}^2=\frac{2}{41}, & b_{\alpha_3}^2=\frac{2}{41}.
\end{array}
\end{equation}
Any permutations of values of $b_\alpha$ are of course also form a solution.

\item

Three dimensional quasi root system number~6 in Appendix~\ref{QR_dim3}.
\begin{equation}
C=
\left(
\begin{array}{ccc}
4 & -1 & -1\\
-1 & 4& 1\\
-1& 1& 4
\end{array}
\right).
\end{equation}
The solutions are
\begin{equation}
                        \label{solution_QR6}
c=\frac{3}{2}
\left(
1\pm\frac{1}{\sqrt{521}}
\right):
\text{ any one of }b_{\alpha}^2 =\frac{88}{1563},  \text{ two other }b_{\alpha}^2=\frac{40}{1563}.
\end{equation}

\item

Three dimensional quasi root system number~4 in Appendix~\ref{QR_dim3}.
\begin{equation}
C=
\left(
\begin{array}{ccc}
4 & -1 & 0\\
-1 & 4& -1\\
0& -1& 4
\end{array}
\right).
\end{equation}
The solutions are
\begin{equation}
                        \label{solution_QR4}
\begin{array}{rlll}
c=\frac{3}{2}\pm \frac{\sqrt{7/103}}{2}: &
b_{\alpha_1}^2 =\frac{9}{206}-\frac{\sqrt{3/2}}{103},
& b_{\alpha_2}^2=\frac{4}{103},
& b_{\alpha_3}^2=\frac{9}{206}+\frac{\sqrt{3/2}}{103}, \\[5pt]
c=\frac{3}{2}\pm \frac{\sqrt{7/103}}{2}: &
b_{\alpha_1}^2 =\frac{9}{206}+\frac{\sqrt{3/2}}{103},
& b_{\alpha_2}^2=\frac{4}{103},
& b_{\alpha_3}^2=\frac{9}{206}-\frac{\sqrt{3/2}}{103},\\
c=\frac{3}{2} \pm  \ii \frac{\sqrt{7/17}}{2}: &
b_{\alpha_1}^2 =\frac{5}{17},
& b_{\alpha_2}^2=\frac{2}{17},
& b_{\alpha_3}^2=\frac{5}{17}.
\end{array}
\end{equation}
The last line represents a complex solution. It is a valid
realization of the Virasoro algebra, but a non-unitary one
-- the $A$ matrix elements are complex in this case.
So formally speaking this solution falls beyond our ansatz.

\end{itemize}

We have obtained a few surprising solutions.
The standard expectation is that the ansatz should
generate solutions of rational central charge only.
But we see that every 3-dimensional $I$-type model has solutions of irrational
central charge. We expect that it is also true for higher dimensional
models, and not only of $I$-type.

The solutions of irrational central charge do not contradict
the ``rationality theorem'' in reference~\cite{Anderson:1987ge}.
The Virasoro field that we obtain in our ansatz~(\ref{main})
is not the energy-momentum tensor field for the theory,
so the corresponding central charge doesn't have to be rational.



\subsection{Solutions for $A$-type root systems}

\label{Solutions for $A$-type root systems}


In this section we want to list the solutions for the models based
on the $A_n$ root systems. These models were solved in~\cite{Neveu:1988fb}
by Neveu and West,
we just rephrase their solution here.

The $A_n$ root systems are $d=n$ dimensional, they consist of $n(n+1)$
long roots. In our normalization the squared length of roots is equal to 4,
and all the mutual products between the roots are even: $0,\pm 2,\pm 4$.
Therefore we can set all the cocycles equal to 1.

To take the advantage of the root system symmetries one has to choose
a special system of coordinates. In the case of $A_n$ root system
one can choose the system of coordinates so that the matrix $A_{\mu \nu}$
is symmetric for all the solutions. In this system of coordinates
the $n$ simple roots look like the following:
the $(i-1)$-th and the $i$-th coordinates of $i$-th simple root are nonzero, the rest
is taken to be zero:
\begin{equation}
                   \label{A_n_simple_roots}
\begin{aligned}
\alpha_1&=\big(2,0,\ldots,0\big),\\
\alpha_2&=\big(-1,\sqrt{3},0,\ldots,0\big),\\
\alpha_3&=\big(0,-\sqrt{4/3},\sqrt{8/3},0,\ldots,0\big),\\
& \cdots\\
\alpha_i&=\big(0,\ldots,0,-\sqrt{2(i-2)/i},\sqrt{2(i+2)/i},0,\ldots,0\big),\\
& \cdots\\
\alpha_n&=\big(0,\ldots,0,-\sqrt{2(n-2)/n},\sqrt{2(n+2)/n}\,\big),
\end{aligned}
\end{equation}
It is easy to see that the vectors above satisfy the desired product
values:
\begin{equation}
(\alpha_i,\alpha_j)=\left\{
\begin{array}{ll}
4, & i=j,\\
-2, & |i-j|=1,\\
0, & |i-j|>1.
\end{array}
\right.
\end{equation}
The rest of positive roots are given by the sums of successive
simple roots: $\alpha_{12}=\alpha_1+\alpha_2,
\alpha_{23}=\alpha_2+\alpha_3, \ldots,
\alpha_{123}=\alpha_1+\alpha_2+\alpha_3,
\alpha_{234}=\alpha_2+\alpha_3+\alpha_4,\ldots,
\alpha_{1\ldots n}=\alpha_1+\cdots +\alpha_n$.
We will denote the coefficients before the vertex operators
in the expression~(\ref{main}) by
$b_1,b_2,\ldots,b_n,b_{12},b_{23},\ldots,b_{123},b_{234},\ldots,
b_{1\ldots n}$. We take the matrix $A$ in~(\ref{main}) to be diagonal
with diagonal elements $a_1,a_2,\ldots,a_n$.

The off-diagonal terms of equation~(\ref{phi_phi_equation})
imply equalities between absolute values of certain coefficients $b$:
\begin{equation}
b_i^2=b_{i-1,i}^2=b_{i-2,i-1,i}^2=\ldots=b_{1,2,\ldots,i-1,i}^2\ ,
\qquad
i=1,\ldots,n,
\end{equation}
so we have only $n$ independent coefficients $b$: $b_1,b_2,\ldots,b_n$.

The detailed solution of the problem can be found in~\cite{Neveu:1988fb}
(section 3, $\mathrm{SU}(N)$ part),
we will not reproduce it here. We only state the result.
It is convenient to label the solutions by a sequence
of $n-1$ numbers $\sigma_i,\ i=1,\ldots,n-1$ that can take values
0 or 1 only. Different sequences represent different solutions.
The $2^{n-1}$ solutions are:
\begin{align}
                    \label{A_n solution a}
a_j&=x_n+\sum_{k=j}^{n-1}x_k \sigma_k (-1)^{\left(\sum_{p=k}^{n-1}\sigma_p\right)},
\quad
\text{where } x_k=\frac{k+1}{k+3}\, ,\\
                    \label{A_n solution c}
c&=\sum_{j=1}^n a_j=
n \, x_n+\sum_{k=1}^{n-1}k\, x_k \sigma_k (-1)^{\left(\sum_{p=k}^{n-1}\sigma_p\right)},\\
                    \label{A_n solution b}
|b_j|&=\left\{
\begin{array}{ll}
\displaystyle \frac{1-a_j}{2}, & \text{if }\displaystyle \left( \sum_{p=j}^{n-1}\sigma_p \right)\text{ even},\\[15pt]
\displaystyle \frac{a_j}{2}, & \text{if } \left(\displaystyle \sum_{p=j}^{n-1}\sigma_p \right) \text{ odd}.
\end{array}
\right.
\end{align}
Note that $a_n=x_n$ and $|b_n|=\frac{1-a_n}{2}$ in all the $2^{n-1}$ solutions.
The other $2^{n-1}$ solutions are obtained using the duality described in the
end of section~\ref{Solutions for the ansatz: general discussion}:
all coefficients $b$ flip the sign, and $a_j \to 1-a_j,\ c\to n-c$.

For brevity we don't specify the signs of $b_j$ here.
One can choose any signs for $b_i, i=1,2,\ldots,n$, then
the signs of the rest of the coefficients $b$ have to be chosen
consistently. 
So the total number of solutions will be $2^{\,2n}$.

Fortunately we obtained all the solutions for the $A_n$ models.
So indeed no solutions exist with the matrix $A$ being non-diagonal.

We want to draw your attention to two types of solutions.
The first type is the symmetric one - it corresponds
to the sequence $\sigma_1=\sigma_2=\ldots=\sigma_{n-1}=0$.
Then all $a_j$ are equal to each other (the matrix $A$ is proportional
to the identity matrix) and
all $|b_j|$ are equal to each other. The two dual symmetric solutions are:
\begin{equation}
\begin{aligned}
c&=n\frac{n+1}{n+3},& a_j&=\frac{n+1}{n+3},& |b_j|&=\frac{1}{n+3},\\
c&=\frac{2\,n}{n+3},& a_j&=\frac{2}{n+3},& |b_j|&=\frac{1}{n+3},
\end{aligned}
\qquad
j=1,2,\ldots,n \, .
\end{equation}
The second solution represents the $\mathbb{Z}_{n+1}$ parafermions~\cite{Fateev:1985mm}.

The second important case is $\sigma_1=\sigma_2=\ldots=\sigma_{n-2}=0$,
$\sigma_{n-1}=1$. Then the central charge in~(\ref{A_n solution c})
becomes
\begin{equation}
c=1-\frac{6}{(n+2)(n+3)},
\end{equation}
which is equal to the central charge of the $n$-th minimal model~\cite{Belavin:1984vu}
of the Virasoro algebra. Therefore the corresponding solution
represents the $n$-th minimal model.

An interesting observation is that all the solutions come in pairs
with the central charge difference equal to $1/2$.
Indeed one can see from equation~(\ref{A_n solution c}) for the
central charge that two solutions labeled by $\sigma$-sequences
with only $\sigma_1$ different and the rest of sigmas identical
have central charge difference $x_1=1/2$. The $a_j, b_j, j>1$ values
are the same for the two solutions, and $a_1$ differ by $1/2$,
$b_1$ differ by $1/4$. We conclude that the Virasoro field expressions
corresponding to the two solutions differ by a field
$L^{\text{Ising}}=-1/4\NO{\pd \phi_1 \pd \phi_1}
\pm 1/4 (\Gamma_{\alpha_1}+\Gamma_{-\alpha_1})$,
representing a solution for the $A_1$ root subsystem.

As an illustration of the general solution for $A_n$ root systems
we would like to give here the explicit numbers for the case $n=3$.
(The $A_2$ case is already solved in section~\ref{$A_2$ root system}.)
The simple roots $\alpha_1$, $\alpha_2$, $\alpha_3$ are as
in~(\ref{A_n_simple_roots}), the other positive roots are
$\alpha_{12}=\alpha_1+\alpha_2$, $\alpha_{23}=\alpha_2+\alpha_3$,
$\alpha_{123}=\alpha_1+\alpha_2+\alpha_3$. The coefficients
of the corresponding vertex operators $\Gamma_\alpha$ are subject
to following equalities:
$|b_3|=|b_{23}|=|b_{123}|, |b_2|=|b_{12}|$.
The relevant numbers $x_k$
in~(\ref{A_n solution a}, \ref{A_n solution c}, \ref{A_n solution b})
are $x_1=1/2,\ x_2=3/5,\ x_3=2/3$.
The solutions are labeled by two binary numbers: $\sigma_1$ and $\sigma_2$.
We present the solutions in a table:
\begin{equation}
\begin{array}{cc|c|ccc|ccc}
\sigma_1 & \sigma_2 & c & a_1 & a_2 & a_3 & |b_1| & |b_2| & |b_3|\\[2pt]
\hline
0 & 0 & 2 & 2/3 & 2/3 & 2/3 & 1/6 & 1/6 & 1/6 \\
\multicolumn{2}{c|}{\mathrm{dual}} & 1 & 1/3 & 1/3 & 1/3 & 1/6 & 1/6 & 1/6 \\
\hline
1 & 0 & 3/2 & 1/6 & 2/3 & 2/3 & 1/12 & 1/6 & 1/6 \\
\multicolumn{2}{c|}{\mathrm{dual}} & 3/2 & 5/6 & 1/3 & 1/3 & 1/12 & 1/6 & 1/6\\
\hline
0 & 1 & 4/5 & 1/15 & 1/15 & 2/3 & 1/30 & 1/30 & 1/6\\
\multicolumn{2}{c|}{\mathrm{dual}} & 11/5 & 14/15 & 14/15 & 1/3 & 1/30 & 1/30 & 1/6\\
\hline
1 & 1 & 13/10 & 17/30 & 1/15 & 2/3 & 13/60 & 1/30 & 1/6\\
\multicolumn{2}{c|}{\mathrm{dual}} & 17/10 & 13/30 & 14/15 & 1/3 & 13/60 & 1/30 & 1/6
\end{array}
\end{equation}


\subsection{Solutions for 3-dimensional quasi root systems}

\label{Solutions for 3-dimensional quasi root systems}


Here we would like to give the solutions for various 3-dimensional
systems, that were not covered in previous sections.

We start from the quasi root system number 7 in the list in Appendix~\ref{QR_dim3}.
It has the smallest number of roots (eight) between all
non-$I$ type 3-dimensional systems. Basically it is $A_2$ root system
plus one more pair of roots, orthogonal to one of the $A_2$ roots.
We choose the coordinates so that
the $A_2$ root system lies in the x-y plane, and the extra root -
in the y-z plane. Then the 4 roots (in accordance with the order chosen
in Appendix~\ref{QR_dim3}) are
\begin{equation}
\begin{aligned}
\alpha_1&=\left(
2,0,0
\right),\\
\alpha_2&=\left(
{-1,\sqrt{3},0}
\right),\\
\alpha_3&=\left(
{0,-\frac{1}{\sqrt{3}},\sqrt{\frac{11}{3}}}
\right),\\
\alpha_4&=\alpha_1+\alpha_2=\left(
{1,\sqrt{3},0}
\right),
\end{aligned}
\end{equation}
The ansatz expression for the Virasoro field is
\begin{equation}
L=-\frac{1}{2}\sum_{\mu,\nu=1}^{3} A_{\mu,\nu} \NO{\pd \phi_\mu \pd \phi_\nu}+
\sum_{i=1}^4 b_i \left( \Gamma_{\alpha_i}+\Gamma_{-\alpha_i} \right).
\end{equation}
Not all the cocycles are equal to 1, but those equal to $-1$ involve
$\alpha_3$ and do not enter to the equations, so we can ignore the cocycles.
We will not write down the 10 equations obtained when one substitutes
the above expression for $L$ to the Virasoro commutation relation.
It is not easy to establish that the equations imply that
either $A_{1,2}=A_{1,3}=0$ or $A_{2,3}=0$. We only list here the final solution,
which was obtained using {\sl Mathematica} on a computer.
The solution is presented in table~\ref{QRSdim3n7}.
For brevity we don't list the dual solutions in the table,
and give only the absolute values of the coefficients $b_i$.
As usual one can choose any sign combination for $b_1, b_2, b_3$
and then the sign of $b_4$ has to be adjusted accordingly.
So the total number of solutions is $6\times 2\times 2^3$,
2 is for dual, $2^3$ counts the sign variations.
\begin{table}[!htb]
\center
$
\begin{array}{|c|c|c|}
\hline
\phantom{a}&&\\[-10pt]
c & A & \bvect{|b_1|}{|b_2|}{|b_3|}{|b_4|}
\\
\phantom{a}&&\\[-10pt]
\hline
\phantom{a}&&\\[-10pt]
\frac{25}{26} &
\left[
\begin{array}{ccc}
 \frac{1}{26} & 0 & 0 \\
 0 & \frac{25}{39} & -\frac{4 \sqrt{11}}{39} \\
 0 & -\frac{4 \sqrt{11}}{39} & \frac{11}{39}
\end{array}
\right]&
\bvect{\frac{1}{52}}{\frac{\sqrt{3}}{13}}{\frac{2}{13}}{\frac{\sqrt{3}}{13}}
\\
\phantom{a}&&\\[-10pt]
\hline
\phantom{a}&&\\[-10pt]
\frac{19}{13}&
\left[
\begin{array}{ccc}
 \frac{7}{13} & 0 & 0 \\
 0 & \frac{25}{39} & -\frac{4 \sqrt{11}}{39} \\
 0 & -\frac{4 \sqrt{11}}{39} & \frac{11}{39}
\end{array}
\right]
&
\bvect{\frac{3}{13}}{\frac{\sqrt{3}}{13}}{\frac{2}{13}}{\frac{\sqrt{3}}{13}}
\\
\phantom{a}&&\\[-10pt]
\hline
\phantom{a}&&\\[-10pt]
\frac{5}{4}-\frac{\sqrt{\frac{11}{179}}}{4} &
\left[
\begin{array}{ccc}
 \frac{1}{4}+\frac{3 \sqrt{\frac{11}{179}}}{4} & 0 & 0 \\
 0 & \frac{1}{2}-\frac{\sqrt{\frac{11}{179}}}{2} &
-\frac{3}{\sqrt{179}} \\
 0 & -\frac{3}{\sqrt{179}} &
\frac{1}{2}-\frac{\sqrt{\frac{11}{179}}}{2}
\end{array}
\right]
&
\bvect{\frac{1+3 \sqrt{\frac{11}{179}}}{8}
}{\sqrt{\frac{5}{179}}
}{\frac{3}{\sqrt{179}}
}{\sqrt{\frac{5}{179}}}
\\
\phantom{a}&&\\[-10pt]
\hline
\phantom{a}&&\\[-10pt]
\frac{7}{4}-\frac{\sqrt{\frac{11}{179}}}{4}
&
\left[
\begin{array}{ccc}
 \frac{3}{4}+\frac{3 \sqrt{\frac{11}{179}}}{4} & 0 & 0 \\
 0 & \frac{1}{2}-\frac{\sqrt{\frac{11}{179}}}{2} &
-\frac{3}{\sqrt{179}} \\
 0 & -\frac{3}{\sqrt{179}} &
\frac{1}{2}-\frac{\sqrt{\frac{11}{179}}}{2}
\end{array}
\right]
&
\bvect{\frac{1-3 \sqrt{\frac{11}{179}}}{8}
}{\sqrt{\frac{5}{179}}
}{\frac{3}{\sqrt{179}}
}{\sqrt{\frac{5}{179}}}
\\
\phantom{a}&&\\[-10pt]
\hline
\phantom{a}&&\\[-10pt]
\frac{3}{2}-\frac{11 \sqrt{\frac{11}{4799}}}{2}
&
\left[
\begin{array}{ccc}
 \frac{1}{2}-\frac{\sqrt{\frac{11}{4799}}}{2} & -\frac{34
\sqrt{\frac{33}{4799}}}{13} & -\frac{133 \sqrt{\frac{3}{4799}}}{13}
\\
 -\frac{34 \sqrt{\frac{33}{4799}}}{13} & \frac{1}{2}-\frac{11 \sqrt{\frac{11}{4799}}}{2} & 0 \\
 -\frac{133 \sqrt{\frac{3}{4799}}}{13} & 0 & \frac{1}{2}+\frac{\sqrt{
\frac{11}{4799}}}{2}
\end{array}
\right]
&
\bvect{
\frac{85 \sqrt{\frac{11}{4799}}}{26}}
{5 \sqrt{\frac{187}{124774}}}
{\frac{3 \sqrt{\frac{4522}{4799}}}{13}}
{\sqrt{\frac{187}{124774}}}
\\
\phantom{a}&&\\[-10pt]
\hline
\phantom{a}&&\\[-10pt]
\frac{3}{2}-\frac{11 \sqrt{\frac{11}{4799}}}{2}
&
\left[
\begin{array}{ccc}
 \frac{1}{2}-\frac{\sqrt{\frac{11}{4799}}}{2} & \frac{34
\sqrt{\frac{33}{4799}}}{13} & \frac{133 \sqrt{\frac{3}{4799}}}{13}
\\
 \frac{34 \sqrt{\frac{33}{4799}}}{13} & \frac{1}{2}-\frac{11 \sqrt{\frac{11}{4799}}}{2} & 0 \\
 \frac{133 \sqrt{\frac{3}{4799}}}{13} & 0 & \frac{1}{2}+\frac{\sqrt{
\frac{11}{4799}}}{2}
\end{array}
\right]
&
\bvect{
\frac{85 \sqrt{\frac{11}{4799}}}{26}}
{ \sqrt{\frac{187}{124774}}}
{\frac{3 \sqrt{\frac{4522}{4799}}}{13}}
{5 \sqrt{\frac{187}{124774}}}
\\[7pt]
\hline
\end{array}
$
                 \caption{Solutions for the 3 dimensional quasi root system number 7.}
                                  \label{QRSdim3n7}
\end{table}


\section{Literature overview}

\label{Literature review}

\setcounter{equation}{0}


In this section we would like to say a few words about the previous works
on the subject and their relation to our paper.
There are essentially just a few papers.


\subsection{Virasoro algebra bosonic constructions}

\label{Virasoro algebra bosonic constructions}


The papers by Corrigan~\cite{Corrigan:gi},
Dunne, Halliday and Suranyi~\cite{Dunne:1988ii},
Neveu and West~\cite{Neveu:1988fb}
use a similar construction:
\begin{equation}
L(z)=
-\frac{1}{2}\sum_{i=1}^d A_{i}\,\NO{\partial \phi_i(z) \partial \phi_i(z)}+
\sum_{
\left( \alpha ,\alpha \right) =4}
 b_{\alpha}\left({\mathrm{e}^{\mathrm{i}\phi_\alpha(z)}}+
{\mathrm{e}^{-\mathrm{i}\phi_\alpha(z)}}\right)
\end{equation}
with some restrictions.
In~\cite{Neveu:1988fb} the above field appears in equation~(3.1),
in~\cite{Dunne:1988ii} -- in equation~(18) (rewritten in different
 notation).
In~\cite{Corrigan:gi} the form of
the Virasoro field is not written explicitly.
In comparison to our ansatz~(\ref{main}) this construction
is less general -- it lacks the terms of type
$\pd \phi\, {\mathrm{e}^{\mathrm{i}\phi_\beta(z)}}$,
where $(\beta,\beta)=2$, and
the matrix $A$ is taken apriori diagonal.
As we know from the examples
 the matrix $A$ can indeed sometimes be chosen diagonal, but in general
 such a choice can (and does) lead to the omission of certain solutions.
Moreover there are additional apriori restrictions
imposed on the systems of roots
and the coefficients in these papers,
which we discuss in detail here.

In~\cite{Corrigan:gi} and in~\cite{Neveu:1988fb}
the construction is based on the ADE-type root systems.
The elements of the twisted group algebra $c_\alpha$ are
 missing in the vertex operators but
 since in the ADE case the root lattice is even all the 2-cocycles $\epsilon$
 can be chosen to be equal to 1, and therefore the cocycles
 can be indeed neglected.

The additional restriction in~\cite{Corrigan:gi} is
that the matrix $A$ is not only diagonal
but proportional to the identity matrix,
and all the coefficients before the vertex operators
are taken to be equal to each over (up to a sign) -- so only symmetric
solutions are obtained. The central charges for symmetric solutions
based on all the ADE-type root systems are obtained in~\cite{Corrigan:gi}.
In~\cite{Neveu:1988fb} the authors consider explicitly
 the cases of $\mathrm{SU}(N)$ and $\mathrm{SO}(2N)$ root systems.
 In the case of $\mathrm{SU}(N)$
 there is a choice of coordinate system 
 that makes $A$ diagonal for all the solutions. The authors fortunately take the correct
 coordinate realization for the root vectors and then obtain
 all the solutions for $\mathrm{SU}(N)$ system (for general $N$),
 and this is the main result of the paper.
 In section~\ref{Solutions for $A$-type root systems}
 we in fact
 take advantage of 
 their remarkable solution.
 However in the case of $\mathrm{SO}(2N)$ root systems the absence of off-diagonal terms
 in matrix $A$ does lead to missed solutions. The authors prove that
 all the solutions for $\mathrm{SO}(2N)$ root systems have integer central charge.
 This is not true of course. It can be seen even from the fact that the
 $\mathrm{SO}(6)$ root system is identical to the $\mathrm{SU}(4)$ root system
 and so should of course have the same set of solutions, but it is not the case
 in~\cite{Neveu:1988fb}.


The paper~\cite{Dunne:1988ii} probably comes closest to our construction.
The authors allow the products between the long roots to be not only even
but also $\pm 1$. They call such systems of vectors ``$\Pi$ systems''.
The authors do not take into account the possibility
of the products $\pm 3$
and don't include
the terms involving vertex operators based on
short vectors in their construction. 
In our terms $\Pi$ systems are quasi root systems which do not include any
short roots.
Since the root lattice is not even anymore the cocycle factors should be included.
The authors do include them in the paper.
The matrix $A$ is still diagonal in their construction.
The authors list all the solutions for the 2-dimensional $\Pi$ systems.
All the six 3-dimensional $\Pi$ systems are listed correctly in Fig.~2
in~\cite{Dunne:1988ii}. A method of construction of $\Pi$ systems
as a product of two ADE-type root systems  is introduced. We
describe this method in section~\ref{Examples of quasi root systems}.

We also would like to mention two papers by Kiritsis \cite{Kiritsis:ny}
and \cite{Kiritsis:1988wr}. In the former the author constructs
a vertex operator realization of the Virasoro algebra,
it is a $c=1/2$ realization which we use as a simplest example, see
expression~(\ref{main_in_one_dimension}).
In~\cite{Kiritsis:1988wr} the author finds the bosonization of
the $\mathbb{Z}_N$ parafermions, i.e.~essentially giving
a symmetric solution for the Virasoro field on $SU(N)$ root system.
In the same paper a more general construction is suggested in a few words:
``it should involve vertex operators based on vectors of square length 4 and mutual
products $0, \pm 1, \pm 2, \pm 3$''. No further details or explicit formulae
are present.


\subsection{$N=2$ superconformal algebra bosonic constructions}

\label{$N=2$ superconformal algebra bosonic constructions}


Another reference which we would like to discuss is
a work by Kazama and Suzuki~\cite{Kazama:1988va}.
The authors study bosonic constructions for the $N=2$
superconformal algebra. Since it contains the Virasoro algebra
as a subalgebra they essentially also obtain
vertex ope\-rator realizations for the Virasoro
algebra as a byproduct. The $N=2$ superconformal algebra is generated by 4 fields:
the Virasoro field $L(z)$, two fermionic fields $G^\pm$ of scaling dimension
$3/2$ and a ``$U(1)$ current'' -- a bosonic field $J$ of scaling dimension
1. The algebraic relations between the fields can be found in formulae~(1-7)
in \cite{Kazama:1988va} or elsewhere. The authors start to build
 vertex operator realizations for the $N=2$ superconformal algebra from
 the following ansatz:
 \begin{equation}
 G^+(z)=\sum_\gamma g(\gamma) {\mathrm{e}^{\mathrm{i}\phi_\gamma(z)}},
 \qquad
 G^-=\sum_\gamma g(\gamma)^\dag {\mathrm{e}^{-\mathrm{i}\phi_\gamma(z)}},
 \end{equation}
 where the coefficients $g(\gamma)$ include the cocycle factors,
 vectors $\gamma$ are of square length 3 and the mutual products between them
 are 0 or 1:
 \begin{equation}
 (\gamma_1,\gamma_2)=\left\{
 \begin{array}{ll}
 3, & \gamma_1=\gamma_2, \\
 0 \text{ or } 1, & \gamma_1 \ne \gamma_2.
 \end{array}
 \right.
 \end{equation}
 So the model data is encoded in the symmetric matrix $\Gamma$ of mutual products
 between vectors $\gamma$:
 \begin{equation}
 \Gamma_{ij}=(\gamma^{(i)},\gamma^{(j)}).
 \end{equation}

 The $U(1)$ current  is expressed then as
 $J(z)=\ii \pd \phi_q(z)$, where $(q,\gamma)=1$ for all $\gamma$.
 The Virasoro field $L(z)$ becomes
 \begin{equation}
           \label{L_KS}
 L=
 -\frac{1}{2}\sum_{\mu,\nu} A_{\mu,\nu} \NO{\pd \phi_\mu \pd \phi_\nu}+
\sum_{\alpha} b_\alpha {\mathrm{e}^{\mathrm{i}\phi_\alpha}},
 \end{equation}
 where
 \begin{equation}
 A_{\mu,\nu}=\frac{1}{2} \sum_{\gamma} |g(\gamma)|^2 \gamma_\mu \gamma_\nu,
 \end{equation}
 the vectors $\alpha$ are square length-4 vectors obtained from the pairs
 of $\gamma$ vectors the product of which is equal to 1: $\alpha=\gamma-\gamma',\
 (\gamma,\gamma')=1$. And the coefficients
 \begin{equation}
 b_\alpha=\frac{1}{2} \sum_
 {\begin{array}{c}
 \scriptstyle \gamma,\gamma'\\[-3pt]
 \scriptstyle(\gamma,\gamma')=1\\[-3pt]
 \scriptstyle \alpha=\gamma-\gamma'
 \end{array}}
 g(\gamma) g(\gamma')^\dag
\end{equation}

Kazama and Suzuki claim that all the operator product expansions
for the $N=2$ superconformal algebra are satisfied providing
the coefficients $g(\gamma)$ obey
\begin{equation}
|g(\gamma^{(i)})|^2=2\sum_j (\Gamma^{-1})_{ij} .
\end{equation}
The central charge of the algebra is then given by
\begin{equation}
c=\frac{3}{2} \sum_\gamma |g(\gamma)|^2=
3 \sum_{ij} (\Gamma^{-1})_{ij}
\end{equation}
The expression for the Virasoro field~(\ref{L_KS}) contains
vertex operators based on long roots only, and it is easy to see
that the products between the different roots can be $0, \pm 1, \pm 2$.
So this Virasoro field should represent a solution for one
of the quasi root systems containing long root only.

Let's give a few examples. The first example is a model based
on matrix
\begin{equation}
\Gamma=
\left(
\begin{array}{ccc}
 3 & 1 & 0 \\
 1 & 3 & 1 \\
 0 & 1 & 3
\end{array}
\right).
\end{equation}
There are three $\gamma$ vectors. They lead to two (not counting the opposite vectors)
$\alpha$ vectors in the expression~(\ref{L_KS}) for the Virasoro
field: $\alpha_{12}=\gamma_1-\gamma_2, \alpha_{23}=\gamma_2-\gamma_3$.
The product between the vectors is equal to $-1$.
So we conclude that vectors $\alpha_{12}, \alpha_{23}$
form  $I_2$ quasi root system. But the $\pd \phi \pd \phi$ part
of the Virasoro field is build from 3 bosons.
(We will not present it here explicitly.)
 The central charge is $15/7$ (one of the unitary minimal models
 of the $N=2$ superconformal algebra). One can easily check that
 the obtained Virasoro field  is just a $c=8/7$ solution~(\ref{I2_c8_7})
 for the $I_2$
 quasi root system plus one free boson.
 It is in fact a general feature of the Kazama-Suzuki models --
 the Virasoro field is the solution for our ansatz based on a
 correspondent quasi root system plus one free boson.

 The second example is a model based
on $\Gamma$ matrix
\begin{equation}
\Gamma=
\left(
\begin{array}{ccc}
 3 & 1 & 1 \\
 1 & 3 & 1 \\
 1 & 1 & 3
\end{array}
\right).
\end{equation}
The central charge here is $c=9/5$, the model represents the third unitary
minimal model of the $N=2$ superconformal algebra.
The Virasoro field in this realization is a symmetric ($c=4/5$)
solution for our ansatz
based on $A_2$ root system plus one free boson.
It's also a general feature of Kazama-Suzuki models
based on $n \times n$ $\Gamma$ matrices with all the off-diagonal elements
equal to 1: it represents the $n$-th unitary minimal model
of the $N=2$ superconformal algebra ($c=\frac{3n}{n+2}$)
and the Virasoro field is a symmetric solution for our ansatz
on $A_{n-1}$ root system plus one free boson.

The data for our last example is given by
$\Gamma$ matrix
\begin{equation}
\Gamma=
\left(
\begin{array}{llll}
 3 & 1 & 0 & 0 \\
 1 & 3 & 1 & 0 \\
 0 & 1 & 3 & 1 \\
 0 & 0 & 1 & 3
\end{array}
\right).
\end{equation}
The model is a $c=30/11$ unitary minimal model. The Virasoro field
in this realization is a sum of that of the free boson field and the $c=19/11$
solution~(\ref{solution_QR5}) for the ansatz based on a three dimensional
quasi root system 
number 5 in Appendix~\ref{QR_dim3}.

Further study of Kazama-Suzuki construction for $N=2$ superconformal algebra
can be found in references~\cite{Cohen:1991ze},\cite{Cohen:1991cc}.


\section{Open problems}

\label{Open problems}

\setcounter{equation}{0}


We didn't touch the representation theory at all in this paper.
However one can also build highest weight representations
of the Virasoro algebra using the vertex operator realization.
Highest weight representations will be realized as combinations
of vertex operators
on the lattice dual to that of the quasi root system used to
construct the Virasoro field.
The description of highest weight representations in this framework
is an interesting subject for the future development.

A classification of quasi root systems is an open problem.
It is very challenging to establish the classification,
but we are even not sure this is feasible at all.
The number of quasi root systems grows very fast with the dimension.
So even if a formal scheme for enumeration of all quasi root systems existed
it would probably be too complex to call it classification.
One could try to restrict the consideration of quasi root systems
to a certain subset using probably physical arguments
coming from conformal field theory. But it is not clear what subset
can be taken and what conformal field theory arguments
can be used. Another possible path on the way to classification
is to modify the notion of simplicity of quasi root systems.
Then hopefully many of the quasi root systems 
can be knocked down to the simple subsystems.
But again we don't know what would be a meaningful
modification of the notion of simplicity.

Another challenge is to solve the coefficient equations in a general way,
i.e.~for a class of quasi root systems and not one by one.
We presented the solutions for the cases of specific quasi root systems,
the only exception is the general solution for the $A_n$ root systems.
We even don't know to prove in general the existence of solutions (in real numbers),
although our experience indicates that solutions always exist
and moreover their number is quite large - grows as $2^d$ with
dimension $d$ of the quasi root system.

In general a classification of quasi root systems combined
with a classification of solutions for the ansatz would
probably provide a classification scheme for unitary models
in two-dimensional conformal field theory,
as we believe that any unitary conformal model
can be bosonized, i.e.~represented in terms of free bosons
and vertex operators.

There could be interesting insights to number theory
(as was suggested by Victor Kac).
It is well known that the Jacobi triple product identity
is connected to the bosonization of free two-dimensional
massless fermion. The identity can be even derived
by computing the characters in two different realizations -
the fermionic one and the bosonic one.
Our bosonization scheme of the Virasoro algebra
could be a source of new number theory identities similar
to the Jacobi triple product.
But this question falls beyond the scope of the present study.


\appendix



\section{More general constructions}

\label{More general constructions}

\setcounter{equation}{0}


We would like to make two comments on our choice of the Virasoro field
in the ansatz~(\ref{main}).

The first comment is that in fact
there is another type of terms of scaling dimension 2
constructed from free bosons, which is missing in our ansatz.
It is the second derivative of a free boson field:
$\pd^2 \phi_\rho(z)$. In conformal field theory the term has already been used
to modify the canonical form of the Virasoro field~(\ref{canonical construction}).
The addition of the second derivative term also shifts the value
of the central charge away from integer.
The construction arises in the framework of the Coulomb gas approach.

It's not very difficult to show that the introduction
of $\pd^2 \phi_\rho$ term doesn't lead to a new theory.
The substitution of the field
\begin{equation}
L(z)=L^\text{ansatz}(z)+ \ii \pd^2 \phi_\rho(z)
\end{equation}
(where $L^\text{ansatz}(z)$ is the ``ansatz field'' from~(\ref{main}))
into Virasoro commutation relation leads again to a number of
quadratic equations for the coefficients.
The equation at the $\pd^2 \phi_\rho$ term reads
$(A \rho)=\rho$. Lets choose a system of coordinates
in which the vector $\rho$ is along the $x$-axis:
$\rho=(|\rho|,0,\ldots,0)$.
Therefore in this coordinates the following components of matrix $A$
should be $A_{11}=1, A_{1i}=0, A_{i1}=0, i=2,\ldots,d$.
The introduction of $\pd^2 \phi_\rho$ term doesn't affect
equation~(\ref{phi_phi_equation}). And since $A_{11}=(A^2)_{11}=1$
the following expression should vanish
\begin{equation}
\sum_{\alpha>0}b_\alpha^2 (\alpha_1)^2 +
2\sum_{\beta>0}\left({(\gamma_\beta)}_1\right)^2 +
\sum_{\beta>0} (\gamma_\beta,\gamma_\beta) (\beta_1)^2
=0.
\end{equation}
This is a sum of squares of real numbers, therefore
every term in the sum should be zero:
\begin{align}
b_\alpha=0 \text{ or } \alpha_1=0,\\
{(\gamma_\beta)}_1=0,\\
(\gamma_\beta,\gamma_\beta)=0 \text{ or } \beta_1=0.
\end{align}
If the coefficients $b_\alpha$ and $\gamma_\beta$ are different
from zero, then all the roots $\alpha, \beta$ and vectors $\gamma$
have vanishing components in the direction of vector $\rho$, so they
belong to a space which is orthogonal to $\rho$. Then obviously
the field $L(z)$ is decomposed to the sum of two ``orthogonal''
fields: one is the standard Coulomb gas expression involving
one free boson in the direction of $\rho$, another is
a field in the form of our ansatz~(\ref{main}) involving
$d-1$  free bosons and vertex operators in the space orthogonal to $\rho$.
Since there is no ``interaction'' between the two fields they
can be studied separately.

The second comment is that we have chosen the coefficients
$b_{\alpha}$, $\gamma_\beta$ to be real. This is of course
 not the most general choice for the vertex operator terms.
If we allow the coefficients to be complex then
the most general unitary expressions for the terms are
\begin{align}
b_{\alpha }\Gamma _{\alpha}(z)&+b_{\alpha }^{\ast }
 \Gamma _{-\alpha }(z),\\
 \mathrm{i} {:}
\partial \phi_{\gamma_\beta}(z)
 \Gamma _{\beta }(z){:}&-
 \mathrm{i} {:}
\partial \phi_{\gamma_\beta^{\ast }}(z)\Gamma _{-\beta }(z){:},
 \end{align}
where $b_{\alpha }$, $\gamma_\beta$ are now a complex number and vector and
$b_{\alpha }^{\ast }$, $\gamma_\beta^{\ast }$ are their complex
conjugates.
In this case the equations for the coefficients
in section~\ref{Solutions for the ansatz: general discussion}
have to be modified a little bit.

Note that one can
use the freedom in the definition of free bosonic fields
to shift them by a constant
$\phi_\mu(z) \to \phi_\mu(z)+\sigma_\mu$
to make $d$ coefficients real.
Then the modified analogues of
equations~(\ref{Gamma_equation}, \ref{Lambda_equation})
usually force the rest of the coefficients
to be real. This is certainly true for the
cases of quasi-root systems that we have solved
in section~\ref{Solutions for the ansatz}.
However we haven't proven in general that one can always choose
all the coefficients to be real.


\section{List of 3-dimensional quasi root systems}

\label{QR_dim3}

\setcounter{equation}{0}


We list below all 15 three-dimensional simple quasi root systems.
There is a set of 3 simple roots in each
three-dimensional simple quasi root system.
One can represent each root in the quasi root system
as a linear combination of simple roots with integer
coefficients, such that all the coefficients are either
non-negative or non-positive.
We would like to emphasize that
the existence of a set of simple roots is not guarantied for a general
quasi root system, but all the three-dimensional quasi root systems
do admit a set of simple roots. Note that this set is usually not unique!

We present the following information for each
three-dimensional simple quasi root system:
the total number of roots, the number of short and long roots separately,
the simple roots (just one choice, remember there can be more than
one choice!), the products between the simple roots, the diagram
representing the products between the simple roots in the graphical form,
and the table of all the roots.

The diagrams are a bit different from the standard Dynkin diagrams.
Filled circles represent long simple roots,
unfilled circles represent short simple roots,
the number of lines between two circles equals
to minus the value of the product between the corresponding roots,
the dashed line represents product value $+1$.

In the table we list the coefficients of each root in its
linear decomposition to simple roots.
Only positive roots are listed, for the negative roots
the coefficients should be reversed.

\begin{enumerate}

\addtolength{\parskip}{-5pt}

\item

$A_3$ root system.\\
Number of roots: 12 (12 long, 0 short).
Generators:
${\alpha _1,\alpha _2,\alpha _3}$.
\\
\\[-7pt]
$
\begin{aligned}
  (\alpha _1,\alpha _2)&=-2, \\
  (\alpha _1,\alpha _3)&=0, \\
  (\alpha _2,\alpha _3)&=-2, \\
\end{aligned}
$
\qquad
\begin{minipage}[c]{52pt}
 \includegraphics[width=50pt]{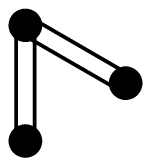}
\end{minipage}
\qquad
$
\begin{array}{l|lll}
  & \alpha _1 & \alpha _2 & \alpha _3 \\
 \hline
 \alpha _1 & 1 & 0 & 0 \\
 \alpha _2 & 0 & 1 & 0 \\
 \alpha _3 & 0 & 0 & 1 \\
 \alpha _4 & 1 & 1 & 0 \\
 \alpha _5 & 0 & 1 & 1 \\
 \alpha _6 & 1 & 1 & 1
\end{array}
$
\\[3pt]


\item

$B_3$ root system. \\
Number of roots: 18 (12 long, 6 short).
Generators:
${\alpha _1,\alpha _2,\beta _1}$.
\\
\\[-7pt]
$
\begin{aligned}
  (\alpha _1,\alpha _2)&=-2, \\
  (\alpha _1,\beta _1)&=0, \\
  (\alpha _2,\beta _1)&=-2, \\
\end{aligned}
$
\qquad
\begin{minipage}[c]{52pt}
 \includegraphics[width=50pt]{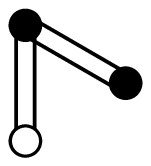}
\end{minipage}
\qquad
$
\begin{array}{l|lll}
  & \alpha _1 & \alpha _2 & \beta _1 \\
 \hline
 \alpha _1 & 1 & 0 & 0 \\
 \alpha _2 & 0 & 1 & 0 \\
 \beta _1 & 0 & 0 & 1 \\
 \alpha _3 & 1 & 1 & 0 \\
 \beta _2 & 0 & 1 & 1 \\
 \beta _3 & 1 & 1 & 1 \\
 \alpha _4 & 0 & 1 & 2 \\
 \alpha _5 & 1 & 1 & 2 \\
 \alpha _6 & 1 & 2 & 2
\end{array}
$
\\[3pt]

\item

$C_3$ root system.\\
Number of roots: 18 (6 long, 12 short).
Generators:
${\alpha _1,\beta _1,\beta _2}$.
\\
\\[-7pt]
$
\begin{aligned}
  (\alpha _1,\beta _1)&=-2, \\
  (\alpha _1,\beta _2)&=0, \\
  (\beta _1,\beta _2)&=-1, \\
\end{aligned}
$
\qquad
\begin{minipage}[c]{52pt}
 \includegraphics[width=50pt]{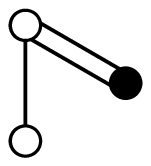}
\end{minipage}
\qquad
$
\begin{array}{l|lll}
  & \alpha _1 & \beta _1 & \beta _2 \\
 \hline
 \alpha _1 & 1 & 0 & 0 \\
 \beta _1 & 0 & 1 & 0 \\
 \beta _2 & 0 & 0 & 1 \\
 \beta _3 & 1 & 1 & 0 \\
 \beta _4 & 0 & 1 & 1 \\
 \beta _5 & 1 & 1 & 1 \\
 \alpha _2 & 1 & 2 & 0 \\
 \beta _6 & 1 & 2 & 1 \\
 \alpha _3 & 1 & 2 & 2
\end{array}
$
\\[3pt]

\newpage
\item

Number of roots: 6 (6 long, 0 short).
Generators:
${\alpha _1,\alpha _2,\alpha _3}$.
\\
\\[-7pt]
$
\begin{aligned}
  (\alpha _1,\alpha _2)&=-1, \\
  (\alpha _1,\alpha _3)&=0, \\
  (\alpha _2,\alpha _3)&=-1, \\
\end{aligned}
$
\qquad
\begin{minipage}[c]{52pt}
 \includegraphics[width=50pt]{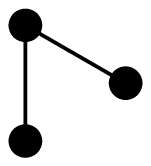}
\end{minipage}
\qquad
$
\begin{array}{l|lll}
  & \alpha _1 & \alpha _2 & \alpha _3 \\
 \hline
 \alpha _1 & 1 & 0 & 0 \\
 \alpha _2 & 0 & 1 & 0 \\
 \alpha _3 & 0 & 0 & 1
\end{array}
$
\\[3pt]

\item

Number of roots: 6 (6 long, 0 short).
Generators:
${\alpha _1,\alpha _2,\alpha _3}$.
\\
\\[-7pt]
$
\begin{aligned}
  (\alpha _1,\alpha _2)&=-1, \\
  (\alpha _1,\alpha _3)&=-1, \\
  (\alpha _2,\alpha _3)&=-1, \\
\end{aligned}
$
\qquad
\begin{minipage}[c]{52pt}
 \includegraphics[width=50pt]{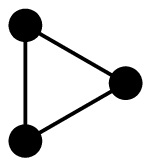}
\end{minipage}
\qquad
$
\begin{array}{l|lll}
  & \alpha _1 & \alpha _2 & \alpha _3 \\
 \hline
 \alpha _1 & 1 & 0 & 0 \\
 \alpha _2 & 0 & 1 & 0 \\
 \alpha _3 & 0 & 0 & 1
\end{array}
$
\\[3pt]

\item

Number of roots: 6 (6 long, 0 short).
Generators:
${\alpha _1,\alpha _2,\alpha _3}$.
\\
\\[-7pt]
$
\begin{aligned}
  (\alpha _1,\alpha _2)&=-1, \\
  (\alpha _1,\alpha _3)&=-1, \\
  (\alpha _2,\alpha _3)&=1, \\
\end{aligned}
$
\qquad
\begin{minipage}[c]{52pt}
 \includegraphics[width=50pt]{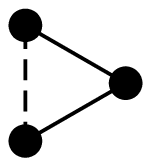}
\end{minipage}
\qquad
$
\begin{array}{l|lll}
  & \alpha _1 & \alpha _2 & \alpha _3 \\
 \hline
 \alpha _1 & 1 & 0 & 0 \\
 \alpha _2 & 0 & 1 & 0 \\
 \alpha _3 & 0 & 0 & 1
\end{array}
$
\\[3pt]

\item

Number of roots: 8 (8 long, 0 short).
Generators:
${\alpha _1,\alpha _2,\alpha _3}$.
\\
\\[-7pt]
$
\begin{aligned}
  (\alpha _1,\alpha _2)&=-2, \\
  (\alpha _1,\alpha _3)&=0, \\
  (\alpha _2,\alpha _3)&=-1, \\
\end{aligned}
$
\qquad
\begin{minipage}[c]{52pt}
 \includegraphics[width=50pt]{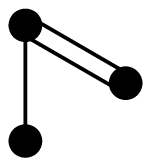}
\end{minipage}
\qquad
$
\begin{array}{l|lll}
  & \alpha _1 & \alpha _2 & \alpha _3 \\
 \hline
 \alpha _1 & 1 & 0 & 0 \\
 \alpha _2 & 0 & 1 & 0 \\
 \alpha _3 & 0 & 0 & 1 \\
 \alpha _4 & 1 & 1 & 0
\end{array}
$
\\[3pt]

\item

Number of roots: 10 (8 long, 2 short).
Generators:
${\alpha _1,\alpha _2,\alpha _3}$.
\\
\\[-7pt]
$
\begin{aligned}
  (\alpha _1,\alpha _2)&=-3, \\
  (\alpha _1,\alpha _3)&=0, \\
  (\alpha _2,\alpha _3)&=-1, \\
\end{aligned}
$
\qquad
\begin{minipage}[c]{52pt}
 \includegraphics[width=50pt]{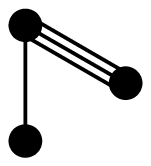}
\end{minipage}
\qquad
$
\begin{array}{l|lll}
  & \alpha _1 & \alpha _2 & \alpha _3 \\
 \hline
 \alpha _1 & 1 & 0 & 0 \\
 \alpha _2 & 0 & 1 & 0 \\
 \alpha _3 & 0 & 0 & 1 \\
 \beta _1 & 1 & 1 & 0 \\
 \alpha _4 & 1 & 1 & 1
\end{array}
$
\\[3pt]

\item

Number of roots: 8 (6 long, 2 short).
Generators:
${\alpha _1,\beta _1,\alpha _2}$.
\\
\\[-7pt]
$
\begin{aligned}
  (\alpha _1,\beta _1)&=-1, \\
  (\alpha _1,\alpha _2)&=-1, \\
  (\beta _1,\alpha _2)&=0, \\
\end{aligned}
$
\qquad
\begin{minipage}[c]{52pt}
 \includegraphics[width=50pt]{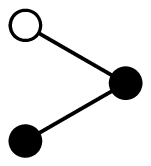}
\end{minipage}
\qquad
$
\begin{array}{l|lll}
  & \alpha _1 & \beta _1 & \alpha _2 \\
 \hline
 \alpha _1 & 1 & 0 & 0 \\
 \beta _1 & 0 & 1 & 0 \\
 \alpha _2 & 0 & 0 & 1 \\
 \alpha _3 & 1 & 1 & 0
\end{array}
$
\\[3pt]

\item

Number of roots: 10 (10 long, 0 short).
Generators:
${\alpha _1,\alpha _2,\alpha _3}$.
\\
\\[-7pt]
$
\begin{aligned}
  (\alpha _1,\alpha _2)&=-2, \\
  (\alpha _1,\alpha _3)&=-1, \\
  (\alpha _2,\alpha _3)&=-1, \\
\end{aligned}
$
\qquad
\begin{minipage}[c]{52pt}
 \includegraphics[width=50pt]{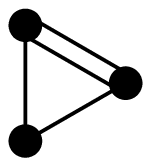}
\end{minipage}
\qquad
$
\begin{array}{l|lll}
  & \alpha _1 & \alpha _2 & \alpha _3 \\
 \hline
 \alpha _1 & 1 & 0 & 0 \\
 \alpha _2 & 0 & 1 & 0 \\
 \alpha _3 & 0 & 0 & 1 \\
 \alpha _4 & 1 & 1 & 0 \\
 \alpha _5 & 1 & 1 & 1
\end{array}
$
\\[3pt]

\item

Number of roots: 12 (6 long, 6 short).
Generators:
${\alpha _1,\beta _1,\beta _2}$.
\\
\\[-7pt]
$
\begin{aligned}
  (\alpha _1,\beta _1)&=-1, \\
  (\alpha _1,\beta _2)&=0, \\
  (\beta _1,\beta _2)&=-1, \\
\end{aligned}
$
\qquad
\begin{minipage}[c]{52pt}
 \includegraphics[width=50pt]{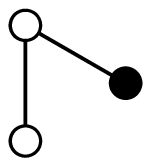}
\end{minipage}
\qquad
$
\begin{array}{l|lll}
  & \alpha _1 & \beta _1 & \beta _2 \\
 \hline
 \alpha _1 & 1 & 0 & 0 \\
 \beta _1 & 0 & 1 & 0 \\
 \beta _2 & 0 & 0 & 1 \\
 \alpha _2 & 1 & 1 & 0 \\
 \beta _3 & 0 & 1 & 1 \\
 \alpha _3 & 1 & 1 & 1
\end{array}
$
\\[3pt]

\item

Number of roots: 12 (8 long, 4 short).
Generators:
${\alpha _1,\alpha _2,\beta _1}$.
\\
\\[-7pt]
$
\begin{aligned}
  (\alpha _1,\alpha _2)&=-1, \\
  (\alpha _1,\beta _1)&=0, \\
  (\alpha _2,\beta _1)&=-2, \\
\end{aligned}
$
\qquad
\begin{minipage}[c]{52pt}
 \includegraphics[width=50pt]{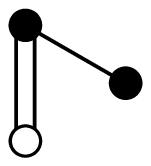}
\end{minipage}
\qquad
$
\begin{array}{l|lll}
  & \alpha _1 & \alpha _2 & \beta _1 \\
 \hline
 \alpha _1 & 1 & 0 & 0 \\
 \alpha _2 & 0 & 1 & 0 \\
 \beta _1 & 0 & 0 & 1 \\
 \beta _2 & 0 & 1 & 1 \\
 \alpha _3 & 1 & 1 & 1 \\
 \alpha _4 & 0 & 1 & 2
\end{array}
$
\\[3pt]

\item

Number of roots: 12 (10 long, 2 short).
Generators:
${\alpha _1,\alpha _2,\beta _1}$.
\\
\\[-7pt]
$
\begin{aligned}
  (\alpha _1,\alpha _2)&=-2, \\
  (\alpha _1,\beta _1)&=0, \\
  (\alpha _2,\beta _1)&=-1, \\
\end{aligned}
$
\qquad
\begin{minipage}[c]{52pt}
 \includegraphics[width=50pt]{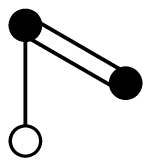}
\end{minipage}
\qquad
$
\begin{array}{l|lll}
  & \alpha _1 & \alpha _2 & \beta _1 \\
 \hline
 \alpha _1 & 1 & 0 & 0 \\
 \alpha _2 & 0 & 1 & 0 \\
 \beta _1 & 0 & 0 & 1 \\
 \alpha _3 & 1 & 1 & 0 \\
 \alpha _4 & 0 & 1 & 1 \\
 \alpha _5 & 1 & 1 & 1
\end{array}
$
\\[3pt]

\newpage
\item

Number of roots: 16 (12 long, 4 short).
Generators:
${\alpha _1,\beta _1,\alpha _2}$.
\\
\\[-7pt]
$
\begin{aligned}
  (\alpha _1,\beta _1)&=-1, \\
  (\alpha _1,\alpha _2)&=0, \\
  (\beta _1,\alpha _2)&=-2, \\
\end{aligned}
$
\qquad
\begin{minipage}[c]{52pt}
 \includegraphics[width=50pt]{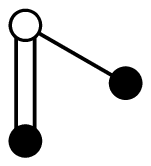}
\end{minipage}
\qquad
$
\begin{array}{l|lll}
  & \alpha _1 & \beta _1 & \alpha _2 \\
 \hline
 \alpha _1 & 1 & 0 & 0 \\
 \beta _1 & 0 & 1 & 0 \\
 \alpha _2 & 0 & 0 & 1 \\
 \alpha _3 & 1 & 1 & 0 \\
 \beta _2 & 0 & 1 & 1 \\
 \alpha _4 & 1 & 1 & 1 \\
 \alpha _5 & 0 & 2 & 1 \\
 \alpha _6 & 1 & 2 & 1
\end{array}
$
\\[3pt]

\item

Number of roots: 20 (12 long, 8 short).
Generators:
${\alpha _1,\beta _1,\beta _2}$.
\\
\\[-7pt]
$
\begin{aligned}
  (\alpha _1,\beta _1)&=-1, \\
  (\alpha _1,\beta _2)&=-1, \\
  (\beta _1,\beta _2)&=-1, \\
\end{aligned}
$
\qquad
\begin{minipage}[c]{52pt}
 \includegraphics[width=50pt]{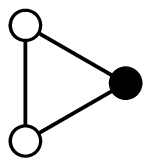}
\end{minipage}
\qquad
$
\begin{array}{l|lll}
  & \alpha _1 & \beta _1 & \beta _2 \\
 \hline
 \alpha _1 & 1 & 0 & 0 \\
 \beta _1 & 0 & 1 & 0 \\
 \beta _2 & 0 & 0 & 1 \\
 \alpha _2 & 1 & 1 & 0 \\
 \alpha _3 & 1 & 0 & 1 \\
 \beta _3 & 0 & 1 & 1 \\
 \beta _4 & 1 & 1 & 1 \\
 \alpha _4 & 1 & 2 & 1 \\
 \alpha _5 & 1 & 1 & 2 \\
 \alpha _6 & 1 & 2 & 2
\end{array}
$
\\[3pt]

\end{enumerate}


\section*{Acknowledgement}


The work on this paper started many years ago.
It was presented on a number of conferences and seminar talks
and discussed with many people (the list would be probably too long).
Especially I am grateful to
Anthony Joseph for many fruitful discussions, interesting ideas
and encouragement, and to Victor Kac for valuable comments
and suggestions.

During the earlier stages of the work my research was supported
by Marie Curie Research Training Network ``LieGrits''
and then by Marie Curie Intra-European Fellowship.
But a substantial part of this paper was written
while I was jobless.
I am very grateful to my wife for her patience and support
during this time.



\end{document}